\documentclass[aps, prd, preprintnumbers, floatfix, nofootinbib, 12pt]{revtex4-2}

\usepackage[dvips]{graphics}
\usepackage{amsmath,amsfonts,amssymb,amsthm}
\usepackage{epsfig,bm}
\usepackage{graphicx,comment}
\usepackage{cancel}
\usepackage{slashed}
\usepackage{verbatim}
\usepackage{latexsym}
\usepackage{float}
\usepackage{enumerate}
\usepackage{array}
\usepackage{tabularx}
\usepackage{longtable}

\usepackage[english]{babel}
\usepackage[utf8]{inputenc}

\usepackage[usenames,dvipsnames]{color}
  \definecolor{hgreen}{rgb}{0,.7,0}
  \definecolor{hred}{rgb}{.7,0,0}
  \definecolor{hblue}{rgb}{0,0,.7}
\usepackage[colorlinks=true,linkcolor=hblue,citecolor=hgreen,filecolor=hblue,urlcolor=hred]{hyperref}

\allowdisplaybreaks

\begin{document}
\title{Flavor Constraints in a Generational Three Higgs Doublet Model}

\author{Wolfgang~Altmannshofer}
\email{waltmann@ucsc.edu}
\affiliation{Department of Physics, University of California Santa Cruz, and
Santa Cruz Institute for Particle Physics, 1156 High St., Santa Cruz, CA 95064, USA}

\author{Kevin~Toner}
\email{ktoner@ucsc.edu}
\affiliation{Department of Physics, University of California Santa Cruz, and
Santa Cruz Institute for Particle Physics, 1156 High St., Santa Cruz, CA 95064, USA}

\begin{abstract}
We propose a Three Higgs Doublet Model (3HDM) that goes beyond natural flavor conservation and in which each of the three Higgs doublets couples mainly to a single generation of fermions via non-standard Yukawa structures. A hierarchy in the vacuum expectation values of the three Higgs doublets can partially address the SM flavor puzzle. In light of the experimentally observed $125$ GeV Higgs boson, we primarily work within a 3HDM alignment limit such that a Standard Model-like Higgs is recovered. In order to reproduce the observed CKM mixing among quarks, the neutral Higgs bosons of the theory necessarily mediate flavor changing neutral currents at the tree level. We consider constraints from neutral kaon, $B$ meson, and $D$ meson mixing as well as from the rare leptonic decays $B_s/B^0/K_L\rightarrow\mu^+\mu^-/e^+e^-$. We identify regions of parameter space in which the new physics Higgs bosons can be as light as a TeV or even lighter.
\end{abstract}
\maketitle

\newpage
\tableofcontents
\newpage

\section{Introduction}
\label{sec:intro}

Measurements of Higgs properties at the LHC~\cite{CMS:2022dwd, ATLAS:2022vkf} indicate that the Higgs is to good approximation Standard Model (SM) like. This is particularly the case for its couplings to gauge bosons and third generation fermions. However, little is still known about the small Higgs couplings to the first and second generation fermions. While evidence for the decay of the Higgs into muons has been established~\cite{CMS:2020xwi, ATLAS:2020fzp}, there is still large room for new physics in all its couplings to the first and second generation.

In this context, it is interesting to speculate that not all fermion masses originate from a single source of electroweak symmetry breaking but that the light fermions might obtain their masses from a second, subdominant source. Setups along this line have been proposed e.g. in~\cite{Altmannshofer:2015esa, Ghosh:2015gpa, Botella:2016krk, Das:1995df, Blechman:2010cs, Egana-Ugrinovic:2019dqu}. Part of the motivation of such scenarios is related to aspects of the SM flavor puzzle~\cite{Altmannshofer:2024jyv}, i.e. the question of why the observed masses of quarks and leptons and the quark mixing exhibit a very hierarchical pattern. The first and second generation of quarks and leptons might be much lighter than the third generation because their masses are proportional to a much smaller source of electroweak symmetry breaking. 

This idea has been implemented in the ``flavorful'' Two Higgs doublet model (2HDM)~\cite{Altmannshofer:2016zrn, Altmannshofer:2017uvs, Altmannshofer:2018bch, Altmannshofer:2019ogm}. One Higgs couples with approximately SM-like interactions to gauge bosons and the third generation fermions, while a second Higgs provides masses for the first and second generation. A mass hierarchy between the third generation and the first two generation can be explained by a hierarchy in vacuum expectation values of the two Higgs bosons.
In this work we explore if this idea can be extended to all three generations in the context of a Three Higgs doublet model (3HDM). 
We are interested in a ``generational'' 3HDM, with each of the three Higgs bosons coupling to a single generation of fermions through rank-1 Yukawa matrices. The observed hierarchy between the masses of the three generations (or part of this hierarchy) may then be explained by a hierarchical pattern of the three Higgs vacuum expectation values. 

A vast literature exists on the topic of 3HDMs, with many studies appearing within the past decade. Aspects of 3HDMs (or in general models with more than two Higgs doublets) that are extensively discussed include the structure of the scalar potential~\cite{Barroso:2006pa, Nishi:2007nh, Ivanov:2010ww, Ivanov:2010wz, Kannike:2012pe, Ivanov:2012ry, Keus:2013hya, Maniatis:2014oza, Ivanov:2014doa, Moretti:2015cwa, Maniatis:2015gma, Pilaftsis:2016erj, Bento:2017eti, Pramanick:2017wry, deMedeirosVarzielas:2019rrp, Das:2019yad, Faro:2019vcd, Darvishi:2019dbh, Ivanov:2020jra, Carrolo:2021euy, Darvishi:2021txa, Kalinowski:2021lvw,  Bento:2022vsb, Boto:2022uwv}, the presence of new sources of CP violation and applications to baryogenesis or leptogenesis~\cite{Weinberg:1976hu, Ahriche:2015mea, deMedeirosVarzielas:2016rii, Ivanov:2018ime, Chakraborty:2019zas, Davoudiasl:2021syn}, the possible presence of a dark matter candidate if at least one the doublets is inert~\cite{Grzadkowski:2009bt, Keus:2014jha, Cordero:2017owj, Cordero-Cid:2018man, Aranda:2019vda, Khater:2021wcx, Kuncinas:2022whn}, interesting collider phenomenology of the additional Higgs bosons~\cite{Akeroyd:2012yg, Merchand:2016ldu, Camargo-Molina:2017klw, Akeroyd:2018axd, Akeroyd:2019mvt, Chakraborti:2021bpy, Ivanov:2021pnr}, as well as other phenomenological implications~\cite{Grimus:2007if, Akeroyd:2016ssd, Solberg:2018aav, Akeroyd:2020nfj, Logan:2020mdz}.

The most relevant aspect for our work is the flavor phenomenology of 3HDMs (see e.g.~\cite{Grossman:1994jb, Cree:2011uy, Hartmann:2014ppa, Yagyu:2016whx, Penuelas:2017ikk, Hernandez:2021iss, Das:2021oik, Boto:2021qgu, Das:2022gbm, Boto:2023nyi}). As we will see, in order to reproduce the observed mixing in the quark sector, our model necessarily violates the principle of natural flavor conservation~\cite{Glashow:1976nt} and therefore features flavor changing neutral currents already at the tree level. The main goal of our paper is to consistently set up the model and to identify the most important flavor constraints. We find that the most generic version of our 3HDM gives large contributions to meson mixing, kaon mixing in particular, from the tree level exchange of neutral Higgs bosons. In such a region of parameter space, the meson mixing constraints push the additional Higgs bosons to scales that are not collider accessible. However, if only the minimal amount of flavor violation required to reproduce the CKM matrix is present, flavor constraints are much more relaxed. In that case, we find that the strongest constraints come from leptonic decays of $B$ mesons and kaons, and Higgs masses around 1-2 TeV are not excluded. An exploration of the distinct collider phenomenology of such a setup is left for future work.

This paper is structured as follows: In section \ref{sec:G3HDM}, we introduce the generational 3HDM. We discuss in detail the scalar sector, including electroweak symmetry breaking, the physical Higgs spectrum, as well as approximations that are valid in the limit of hierarchical vacuum expectation values. We also spell out our assumptions about the Yukawa sector and determine the couplings of the physical Higgs bosons with quarks and charged leptons. In section \ref{sec:flavorprobe}, we discuss the most relevant low-energy flavor constraints of our model. We cover neutral meson mixing and rare leptonic decays of $B$ mesons and kaons, $B^0 \to \ell^+ \ell^-$, $B_s \to \ell^+ \ell^-$, and $K \to \ell^+\ell^-$. We conclude and summarize our results in section~\ref{sec:conclusions}. Details about renormalization group running for meson mixing constraints and a simple setup that generates rank-1 Yukawa couplings are presented in appendices~\ref{appendix:mix} and~\ref{app:rank1}, respectively. 

\section{A generational three Higgs doublet model}
\label{sec:G3HDM}

\subsection{The field content and the Lagrangian} \label{sec:lagrangian}

We augment the SM with two additional Higgs doublets.
The associated gauge representations under $SU(3)_c \times SU(2)_L \times U(1)_Y$ are analogous to the SM. All three Higgs doublets transform as $\Phi_a \sim ({\mathbf 1},{\mathbf 2},\frac{1}{2})$, with $a = 1,2,3$ labeling the three Higgs fields. 

The terms in the model's Lagrangian that contain the Higgs fields can be written as
\begin{equation}
    \mathcal{L}_\text{3HDM} \supset
    \sum \limits_{a=1}^{3}
    \lvert D_{\mu} \Phi_a \rvert ^2 - V_\text{3HDM} + \mathcal{L}_\text{3HDM}^\text{Yuk}
    ~.
\end{equation}
In addition to the kinetic terms, the Lagrangian contains a potential for the Higgs fields and Yukawa interactions with the SM fermions.
The most general Yukawa Lagrangian is
\begin{multline}
\label{eq:lag}
    -\mathcal{L}_\text{3HDM}^\text{Yuk} = \sum\limits_{i,j} \left( \lambda_{u_1}^{ij} \bar{q}_{L_i} \tilde{\Phi}_1 u_{R_j} + \lambda_{d_1}^{ij}\bar{q}_{L_i} \Phi_1 d_{R_j}+\lambda_{\ell_1}^{ij} \bar{\ell}_{L_i} \Phi_1 e_{R_j} \right) + \text{h.c.}
    \\
    + \sum\limits_{i,j} \left( \lambda_{u_2}^{ij} \bar{q}_{L_i} \tilde{\Phi}_2 u_{R_j} + \lambda_{d_2}^{ij} \bar{q}_{L_i} \Phi_2 d_{R_j} + \lambda_{\ell_2}^{ij} \bar{\ell}_{L_i} \Phi_2 e_{R_j} \right) + \text{h.c.}
    \\
    + \sum\limits_{i,j} \left( \lambda_{u_3}^{ij} \bar{q}_{L_i} \tilde{\Phi}_3 u_{R_j} + \lambda_{d_3}^{ij} \bar{q}_{L_i} \Phi_3 d_{R_j} + \lambda_{\ell_3}^{ij} \bar{\ell}_{L_i} \Phi_3 e_{R_j} \right) + \text{h.c.}~,
\end{multline}
where $\Tilde{\Phi}_{a} \equiv i\sigma_{2} \Phi_{a}^{*}$ and $i,j$ are flavor indices which run from 1 to 3. $\overline{q}_L$ and $\overline{\ell}_L$ represent the left-handed quark and lepton doublets, while $u_R$, $d_R$, $e_R$ are the quark and lepton singlets. The $\lambda$ matrices denote the Yukawa couplings. Neutrino masses and mixing are beyond the scope of this paper.

The most general gauge invariant $n$HDM potential may contain up to $n^2$ mass parameters and $\frac{n^2}{2}(n^2+1)$ quartic interactions (see e.g. \cite{Bento:2017eti, Darvishi:2019dbh})
\begin{equation}
    V_{\text{3HDM}}
    =
    \sum\limits_{a,b=1}^{3}
    m_{ab}^2
    (\Phi_{a}^{\dag}\Phi_{b})
    +
    \sum\limits_{a,b,c,d=1}^{3}
    \lambda_{ab,cd}
    (\Phi_{a}^{\dag}\Phi_{b})
    (\Phi_{c}^{\dag}\Phi_{d})
    ~,
\end{equation}
where $\lambda_{ab,cd}=\lambda_{cd,ab}$ and, by hermiticity, $m_{ab}^2=(m_{ba}^2)^*$ while $\lambda_{ab,cd}=\lambda^{*}_{ba,dc}$.
In a 3HDM, this corresponds in general to 54 terms in the Higgs potential.

As discussed in more detail in section~\ref{sec:yukawa} and appendix~\ref{app:rank1}, we are interested in a scenario in which the three Higgs fields couple preferentially to a single generation. This can be most readily achieved if the Higgs fields are charged under softly broken $U(1)$ flavor symmetries. We thus consider it plausible that the Higgs potential respects to a good approximation a softly broken $U(1)^3$ symmetry, with the three $U(1)$ factors acting separately on a single Higgs field. This reduces significantly the possible 54 potential terms.
Explicitly, we are left with\footnote{This setup is sometimes referred to as the $U(1) \times U(1)$ symmetric 3HDM in the literature, see e.g.~\cite{Faro:2019vcd}. In fact, the third $U(1)$ is automatically guaranteed by hypercharge gauge invariance.}
\newpage
\begin{align}
\label{eq:V3HDM}
    V_\text{3HDM} &= m_{11}^2(\Phi_{1}^{\dag}\Phi_{1}) + m_{22}^2(\Phi_{2}^{\dag}\Phi_{2}) + m_{33}^2(\Phi_{3}^{\dag}\Phi_{3}) \nonumber
    \\
    &~~ -
    \left[
    m_{12}^2(\Phi_{1}^{\dag}\Phi_{2}) + m_{23}^2(\Phi_{2}^{\dag}\Phi_{3})
    + m_{13}^2(\Phi_{1}^{\dag}\Phi_{3})
    + \text{h.c.}
    \right] \nonumber
    \\
    &~~ +
    \lambda_{1}(\Phi_{1}^{\dag}\Phi_{1})^2
    +
    \lambda_{2}(\Phi_{2}^{\dag}\Phi_{2})^2
    +
    \lambda_{3}(\Phi_{3}^{\dag}\Phi_{3})^2
    +
    \lambda_{4}(\Phi_{1}^{\dag}\Phi_{1})(\Phi_{2}^{\dag}\Phi_{2})
    +
    \lambda_{5}(\Phi_{1}^{\dag}\Phi_{1})(\Phi_{3}^{\dag}\Phi_{3}) \nonumber
    \\
    &~~ +
    \lambda_{6}(\Phi_{2}^{\dag}\Phi_{2})(\Phi_{3}^{\dag}\Phi_{3})
    +
    \lambda_{7}(\Phi_{1}^{\dag}\Phi_{2})(\Phi_{2}^{\dag}\Phi_{1})
    +
    \lambda_{8}(\Phi_{1}^{\dag}\Phi_{3})(\Phi_{3}^{\dag}\Phi_{1})
    +
    \lambda_{9}(\Phi_{2}^{\dag}\Phi_{3})(\Phi_{3}^{\dag}\Phi_{2}) 
    ~.
\end{align}
The three diagonal mass parameters $m_{aa}^2$ as well as all nine quartic interactions $\lambda_i$ of the potential are real. The three off-diagonal mass parameters, $m^2_{ab}$ with $a \neq b$, softly break the $U(1)^3$ symmetry. They can be complex and give rise to CP violation. Note that one can in principle use the freedom to re-phase the Higgs fields and remove the imaginary parts of two of the off-diagonal mass parameters, leaving a single physical CP violating phase.

The quartic couplings are not fully arbitrary but are constrained by bounded from below and vacuum stability conditions (see e.g.~\cite{Barroso:2006pa, Kannike:2012pe, Faro:2019vcd, Boto:2022uwv}), as well as by perturbativity considerations (see e.g.~\cite{Moretti:2015cwa}). 

\subsection{Electroweak symmetry breaking and the Higgs spectrum}

We assume that the three Higgs fields acquire vacuum expectation values such that spontaneous symmetry breaking (SSB) occurs: $SU(2)_L \times U(1)_Y \rightarrow U(1)_{\text{em}}$.
In particular, we assume that the vacuum expectation values (vevs) of the Higgs fields are aligned such that $U(1)_{\text{em}}$ remains unbroken.

Based on the study of bounded from below constraints in 3HDMs that include charged runaway directions (see e.g.~\cite{Faro:2019vcd, Boto:2022uwv}) and the study of electroweak symmetry breaking in multi-Higgs doublet models (see e.g.~\cite{Nishi:2007nh, Maniatis:2014oza}), we expect that an order 1 amount of parameter space does not break $U(1)_\text{em}$, as is common in the 3HDM literature.

In this case, we can parameterize the scalar doublets in terms of charged and neutral components in the usual way
\begin{equation}
    \Phi_a=
    \begin{pmatrix}
    \phi^{+}_a \\ \frac{1}{\sqrt{2}} \left( v_a+\varphi_a+ia_a \right)
    \end{pmatrix}
    ~.
\end{equation}
The three vevs $v_a$ can, in principle, be complex. We find it convenient to work in a phase convention in which all three vevs are real. 
Without loss of generality, we label the Higgs fields such that $v_1 < v_2 < v_3$. As discussed below in section~\ref{sec:yukawa}, we will construct the Yukawa sector such that the doublets $\Phi_1$, $\Phi_2$, and $\Phi_3$ provide mass mainly to the first, second, and third generation of fermions, respectively. The ordering of the three vevs thus follows the mass ordering of the three generations of fermions, and we will later focus mainly on the limit $v_1 \ll v_2 \ll v_3$, which partially addresses the hierarchies in the fermion spectrum of the SM.

In order to reproduce the masses of the $W$ and $Z$ bosons, the individual vevs must sum in quadrature to the SM Higgs vev
\begin{equation}
    v_{\text{SM}}^{2} \equiv v^2 = v_1^2 + v_2^2 + v_3^2 \simeq (246~\text{GeV})^{2} 
    ~.
\end{equation}
This condition is satisfied by construction if we work with the convenient parameterization
\begin{equation} \label{eq:beta_betap}
    v_1=v\,\cos\beta^\prime = v c_{\beta^\prime}
    ~,~~~
    v_2=v\,\sin\beta^\prime\,\cos\beta = v s_{\beta^\prime} c_\beta
    ~,~~~
    v_3=v\,\sin\beta^\prime\,\sin\beta = v s_{\beta^\prime} s_\beta
    ~,
\end{equation}
where we used the notation $s_x = \sin x$, and $c_x = \cos x$. The above definitions imply
\begin{equation}
    \tan\beta^\prime = t_{\beta^\prime} = \frac{\sqrt{v_2^2+v_3^2}}{v_1} ~,~~~
    \tan\beta= t_\beta = \frac{v_3}{v_2}~,
\end{equation}
which may be viewed as an extension of the 2HDM definition for $\tan\beta$. In effect, we may trade $v_1$, $v_2$, $v_3$ for $v$, $\tan{\beta}$, $\tan{\beta^\prime}$.

Working with real vevs implies in general that $m^2_{12}$, $m^2_{13}$, and $m^2_{23}$ all have imaginary parts. We find that these imaginary parts are related by the minimization conditions of the potential in the following way
\begin{equation}
 \text{Im}(m_{13}^2) = - \frac{v_2}{v_3}~ \text{Im}(m_{12}^2) ~,\qquad  \text{Im}(m_{23}^2) = \frac{v_1}{v_3}~ \text{Im}(m_{12}^2) ~.
\end{equation}
The remaining minimization conditions allow us to express the mass parameters $m_{11}^2$, $m_{22}^2$, and $m_{33}^2$ in terms of the real vacuum expectation values
\begin{align} \label{eq:min1}
    m^2_{11} &= \text{Re}(m^2_{12}) \frac{v_2}{v_1}+ \text{Re}(m^2_{13}) \frac{v_3}{v_1}-\lambda^2_{1}v^2_{1}-\frac{1}{2}\big[(\lambda_4 +\lambda_7)v_2^2+(\lambda_5 +\lambda_8)v_3^2\big]
    ~,
    \\  \label{eq:min2}
    m^2_{22} &= \text{Re}(m^2_{12}) \frac{v_1}{v_2}+ \text{Re}(m^2_{23}) \frac{v_3}{v_2}-\lambda^2_{2}v^2_{2}-\frac{1}{2}\big[(\lambda_4 +\lambda_7)v_1^2+(\lambda_6 +\lambda_9)v_3^2\big]
    ~,
    \\  \label{eq:min3}
    m^2_{33} &= \text{Re}(m^2_{13}) \frac{v_1}{v_3}+ \text{Re}(m^2_{23}) \frac{v_2}{v_3}-\lambda^2_{3}v^2_{3}-\frac{1}{2}\big[(\lambda_5 +\lambda_8)v_1^2+(\lambda_6 +\lambda_9)v_2^2\big]
    ~.
\end{align}
In what follows, we will make the simplifying assumption that the Higgs potential respects CP invariance and set $\text{Im}(m_{12}^2) = \text{Im}(m_{13}^2) = \text{Im}(m_{23}^2) =0$. The general case with CP violation in the Higgs potential and the possible implications will be discussed elsewhere. 

In the absence of CP violation, $n$HDM models will contain $n$ physical neutral CP-even Higgs bosons, $n-1$ physical neutral CP-odd Higgs bosons, and $2(n-1)$ physical charged Higgs bosons. The remaining CP-odd and charged degrees of freedom are Goldstone bosons ($G^0, G^\pm$) that provide the longitudinal components of the $Z$ and $W$ bosons of the SM. In our 3HDM case, after SSB, we expect a total of 3 neutral CP-even Higgs bosons ($h$, $H$, $H^\prime$), 2 neutral CP-odd Higgs ($A$, $A^\prime$), and 2 pairs of charged Higgs bosons ($H^\pm$, $H^{\pm \, \prime}$). 

The mass matrix of the CP-odd scalars is independent of the quartic interactions and given by 
\begin{equation}
    \hat{m}^{2}_{a}=
    \begin{pmatrix}
    m_{12}^2~\frac{v_{2}}{v_{1}}+m_{13}^2~\frac{v_{3}}{v_{1}} & -m_{12}^2 & -m_{13}^2\\
    -m_{12}^2 & m_{12}^2~\frac{v_{1}}{v_{2}}+m_{23}^2~\frac{v_{3}}{v_{2}} & -m_{23}^2\\
    -m_{13}^2 & -m_{23}^2 & m_{13}^2~\frac{v_{1}}{v_{3}}+m_{23}^2~\frac{v_{2}}{v_{3}}
    \end{pmatrix}
    ~.
\end{equation}
For the charged and CP-even scalar mass matrices, we find 
\begin{eqnarray}
    \hat{m}^{2}_{\pm} &=&
    \hat{m}^{2}_{a} + \frac{1}{2}
    \begin{pmatrix}
    -v_2^2 \lambda_7 - v_3^2 \lambda_8 & v_1 v_2 \lambda_7 & v_1 v_3 \lambda_8\\
    v_1 v_2 \lambda_7 & -v_1^2 \lambda_7 - v_3^2 \lambda_9 & v_2 v_3 \lambda_9\\
    v_1 v_3 \lambda_8 & v_2 v_3 \lambda_9 & -v_1^2 \lambda_8 - v_2^2 \lambda_9)
    \end{pmatrix}
    ~,
    \\
    \hat{m}^{2}_{\varphi} &=& \hat{m}^{2}_{a}+
    \begin{pmatrix}
    2\,v_1^2 \lambda_1 & v_1 v_2 (\lambda_4 +\lambda_7) & v_1 v_3 (\lambda_5 +\lambda_8)\\
    v_1 v_2 (\lambda_4 +\lambda_7) & 2\,v_2^2 \lambda_2 & v_2 v_3 (\lambda_6 +\lambda_9)\\
    v_1 v_3 (\lambda_5 +\lambda_8) & v_2 v_3 (\lambda_6 +\lambda_9) & 2\,v_3^2 \lambda_3
    \end{pmatrix}
    ~.
\end{eqnarray}

We rotate into the physical Higgs mass-eigenstate basis through orthogonal, 3-parameter rotations for each of the three Higgs sectors 
\begin{equation} \label{eq:higgs_rotations}
    \begin{pmatrix}
    A^\prime \\ A \\ G^0
    \end{pmatrix} = O_{A}
    \begin{pmatrix}
    a_1 \\ a_2 \\ a_3
    \end{pmatrix}
    ~,
    \qquad
    \begin{pmatrix}
    H^{\pm\prime} \\ H^{\pm} \\ G^{\pm}
    \end{pmatrix} = O_{\pm}
    \begin{pmatrix}
    \phi_{1}^{\pm} \\
    \phi_{2}^{\pm} \\    
    \phi_{3}^{\pm}
    \end{pmatrix}
    ~,
    \qquad
    \begin{pmatrix}
    H^\prime \\ H \\ h
    \end{pmatrix} = O_{H}
    \begin{pmatrix}
    \varphi_1 \\
    \varphi_2 \\    
    \varphi_3
    \end{pmatrix} 
    ~.
\end{equation}
The diagonalization matrices $O_A$, $O_\pm$, and $O_H$ can be parameterized as products of three rotation matrices. We find for the pseudoscalar and charged Higgs sectors
\begin{equation}
O_A = \begin{pmatrix}  \label{eq:OA}
    \cos\gamma & -\sin\gamma & 0\\
    \sin\gamma & \cos\gamma & 0\\
    0 & 0 & 1
    \end{pmatrix}\begin{pmatrix}
    \sin{\beta^\prime} & 0 & -\cos{\beta^\prime} \\
    0 & 1 & 0\\
    \cos{\beta^\prime} & 0 & \sin{\beta^\prime}
    \end{pmatrix}\begin{pmatrix}
    1 & 0 & 0\\
    0 & \sin\beta & -\cos\beta\\
    0 & \cos\beta & \sin\beta
    \end{pmatrix} ~,
\end{equation}
\begin{equation} \label{eq:Opm}
    O_\pm = \begin{pmatrix}
    \cos{\gamma_\pm} & -\sin{\gamma_\pm} & 0\\
    \sin{\gamma_\pm} & \cos{\gamma_\pm} & 0\\
    0 & 0 & 1
    \end{pmatrix}\begin{pmatrix}
    \sin{\beta^\prime} & 0 & -\cos{\beta^\prime} \\
    0 & 1 & 0\\
    \cos{\beta^\prime} & 0 & \sin{\beta^\prime}
    \end{pmatrix}\begin{pmatrix}
    1 & 0 & 0\\
    0 & \sin\beta & -\cos\beta\\
    0 & \cos\beta & \sin\beta
    \end{pmatrix} ~,
\end{equation}
where $\beta$ and $\beta^\prime$ where already introduced in \eqref{eq:beta_betap}. 
After applying the $\beta$ and $\beta^\prime$ rotations, one obtains partially diagonalized forms of the respective mass matrices, with the massless Goldstone bosons $G^0$ and $G^\pm$ already appearing as eigenstates. The final rotations by the angels $\gamma$ and $\gamma_\pm$ fully diagonalize the mass matrices and define the massive eigenstates $A$, $A^\prime$ and $H^\pm$, $H^{\pm \, \prime}$.

The pseudoscalar masses $m_A$, $m_{A^\prime}$ and the mixing angle $\gamma$ are determined by
\begin{eqnarray}
  m_A^2 m_{A^\prime}^2 &=& \frac{m_{23}^2}{c_{\beta^\prime} s_{\beta^\prime}} \left( \frac{m_{13}^2}{c_\beta} + \frac{m_{12}^2}{s_\beta} \right) + \frac{m_{12}^2 m_{13}^2}{c_\beta s_\beta s_{\beta^\prime}^2} ~, \\
  m_A^2 + m_{A^\prime}^2 &=& \frac{m_{23}^2}{c_\beta s_\beta} + \frac{1}{c_{\beta^\prime} s_{\beta^\prime}} \left[ m_{12}^2 \left(\frac{s_\beta^2 c_{\beta^\prime}^2}{c_\beta} + c_\beta\right) + m_{13}^2 \left(\frac{c_\beta^2 c_{\beta^\prime}^2}{s_\beta} + s_\beta\right)  \right]~, \\
 (m_A^2 - m_{A^\prime}^2) \frac{1}{2}\sin(2 \gamma) &=& m_{13}^2 \frac{c_\beta}{s_{\beta^\prime}} - m_{12}^2 \frac{s_\beta}{s_{\beta^\prime}} ~.
\end{eqnarray}
These relations can be used to trade the Lagrangian parameters $m_{12}^2$, $m_{13}^2$, $m_{23}^2$ for $m_A$, $m_{A^\prime}$, $\gamma$. The charged Higgs masses $m_{H^\pm}$, $m_{H^{\pm \, \prime}}$ and the corresponding mixing angle $\gamma_\pm$ can be determined analogously. We find
\begin{eqnarray}
  m_{H^\pm}^2 m_{H^{\pm \,\prime}}^2 &=& m_A^2 m_{A^\prime}^2 - \frac{v^2}{2} \Bigg[ \frac{\lambda_7}{s_\beta} \left( m_{23}^2 c_\beta + m_{13}^2 / t_{\beta^\prime} \right) \nonumber \\
  && ~~ + \frac{\lambda_8}{c_\beta} \left( m_{23}^2 s_\beta + m_{12}^2 / t_{\beta^\prime} \right) + \lambda_9 t_{\beta^\prime} \left( m_{12}^2 c_\beta + m_{13}^2 s_\beta \right) \Bigg] \nonumber \\
  && ~~ + \frac{v^4}{4} \Bigg[ \lambda_7 \lambda_8 c_{\beta^\prime}^2 +\lambda_7 \lambda_9 c_\beta^2 s_{\beta^\prime}^2 + \lambda_8 \lambda_9 s_\beta^2 s_{\beta^\prime}^2 \Bigg]  ~, \\
  m_{H^\pm}^2 + m_{H^{\pm \, \prime}}^2 &=& m_A^2 + m_{A^\prime}^2 -\frac{v^2}{2} \Bigg[ \lambda_7 \left(c_\beta^2 + s_\beta^2 c_{\beta^\prime}^2 \right) + \lambda_8 \left( s_\beta^2 + c_\beta^2 c_{\beta^\prime}^2 \right) + \lambda_9 s_{\beta^{\prime}}^2 \Bigg]~, \\
 (m_{H^\pm}^2 - m_{H^{\pm \, \prime}}^2) \frac{1}{2}\sin(2 \gamma_\pm) &=& (m_A^2 - m_{A^\prime}^2) \frac{1}{2}\sin(2 \gamma) + \frac{v^2}{2} \left(\lambda_7-\lambda_8 \right)  c_\beta s_\beta c_{\beta^\prime} ~.
\end{eqnarray}

In the case of the CP-even Higgs bosons, we write
\begin{equation} \label{eq:OH}
    O_H = \begin{pmatrix}
    \cos{\gamma_H} & -\sin{\gamma_H} & 0\\
    \sin{\gamma_H} & \cos{\gamma_H} & 0\\
    0 & 0 & 1
    \end{pmatrix}\begin{pmatrix}
    \cos{\alpha^\prime} & 0 & \sin{\alpha^\prime} \\
    0 & 1 & 0\\
    -\sin{\alpha^\prime} & 0 & \cos{\alpha^\prime}
    \end{pmatrix}\begin{pmatrix}
    1 & 0 & 0\\
    0 & \cos\alpha & \sin\alpha\\
    0 & -\sin\alpha & \cos\alpha
    \end{pmatrix} ~,
\end{equation}
with all three angles $\alpha$, $\alpha^\prime$, and $\gamma_H$, as new parameters.
As already anticipated in the introduction, the stringent limits from meson mixing generically suggest that the BSM Higgs bosons may be considerably heavier than the electroweak scale. It is thus motivated to consider the decoupling limit in which $v_1^2, v_2^2, v_3^2 \ll m_{12}^2, m_{13}^2, m_{23}^2$. In this limit, one finds to first approximation 
\begin{equation} \label{eq:decoupling_LO}
 m_H^2 \simeq m_A^2 ~,~~~ m_{H^\prime}^2 \simeq m_{A^\prime}^2~,~~~ O_H \simeq O_A ~.
\end{equation}
The eigenstate $h$ has a mass of the order of $v$ and has at leading order precisely SM-like couplings. It is thus identified with the 125~GeV Higgs. As we will see in section~\ref{sec:flavorprobe}, in the approximation~\eqref{eq:decoupling_LO}, an exact cancellation of new physics contributions to the meson mixing amplitudes can occur. To assess the constraints from meson mixing, it thus becomes important to systematically include corrections beyond the leading order of the decoupling limit.

\subsection{Beyond the leading order of the decoupling limit with large \texorpdfstring{\boldmath$\tan\beta$}{tan(beta)} and \texorpdfstring{\boldmath$\tan\beta^\prime$}{tan(beta')}} \label{sec:beyond_decoupling_limit}

While it is possible to systematically expand the masses and mixing angles in the CP even scalar sector in powers of $v_k^2/m_{ij}^2$, the resulting expressions are rather lengthy and will not be presented here. Instead, we will focus on the scenario of large $\tan\beta$ and $\tan\beta^\prime$ in which the expressions simplify considerably. In fact, one motivation to consider the generational 3HDM model is the possibility of partially addressing the hierarchies in the SM fermion masses. We thus consider a scenario with $v_1 \ll v_2 \ll v_3$, corresponding to $1 \ll \tan\beta \ll \tan\beta^\prime$.

First of all, assuming no particular hierarchy in the mass parameters $m_{12}^2 \sim m_{13}^2 \sim m_{23}^2$, we find that the pseudoscalar $A^\prime$ is parametrically heavier than $A$ \footnote{According to the minimization conditions in eqs.~\eqref{eq:min1} - \eqref{eq:min3}, this corresponds to a hierarchical set of diagonal masses $m^2_{11} \gg m^2_{22} \gg m^2_{33} \sim O(v^2)$. Alternatively, one could entertain a scenario with $m^2_{11} \sim m^2_{22} \gg m^2_{33} \sim O(v^2)$, which gives $m_{A^\prime}^2 \sim m_A^2$, but requires $m_{13}^2 \ll m_{23}^2$. In our numerical analysis, we do not assume any particular hierarchy in the masses of the heavy Higgs bosons.}
\begin{equation}
    m_{A^\prime}^2 \simeq m_{13}^2 \tan\beta^\prime \gg m_A^2 \simeq  m_{23}^2 \tan\beta ~. 
\end{equation}
Moreover, the mixing angle in the pseudoscalar sector, $\gamma$, turns out to be small and is of the order of $\gamma \sim 1/\tan\beta^\prime \ll 1$. More precisely, we find 
\begin{eqnarray}
    \gamma_A \simeq \frac{m_{12}^2}{m_{13}^2} \, \frac{1}{\tan\beta^\prime} ~.
\end{eqnarray}

In the charged Higgs sector, the masses and mixing angles are given at leading order by the corresponding pseudoscalar quantities. Including next-to-leading order corrections, we find
\begin{equation}
m_{H^\pm}^2 \simeq m_A^2 - \frac{v^2}{2} \lambda_9 ~,~~~ m_{H^{\pm\,\prime}}^2 \simeq m_{A^\prime}^2 - \frac{v^2}{2} \lambda_8 ~,~~~ \gamma_\pm \simeq \gamma  + \frac{\gamma v^2}{2m_{A^\prime}^2} (\lambda_8 - \lambda_9 ) ~.
\end{equation}
Similarly, for the masses of the scalar Higgs bosons we find
\begin{eqnarray}
m_h^2 \simeq 2 v^2 \lambda_3 ~,~~~ m_{H}^2 &\simeq& m_A^2 + \frac{2v^2}{\tan^2\beta} \left(\lambda_2 +\lambda_3 - \lambda_6 - \lambda_9\right) = m_A^2 + \frac{2v^2}{\tan^2\beta} \lambda_H ~, \\
m_{H^\prime}^2 &\simeq& m_{A^\prime}^2 + \frac{2v^2}{\tan^2\beta^\prime} \left(\lambda_1 +\lambda_3 - \lambda_5 - \lambda_8\right) = m_{A^\prime}^2 + \frac{2v^2}{\tan^2\beta^\prime} \lambda_{H^\prime}~.
\end{eqnarray}
For the diagonalization matrix in the scalar Higgs sector, we find it convenient to write
\begin{equation}
    O_H = (1\!\!1 + \Delta) O_A ~,\quad \Delta \simeq \begin{pmatrix}
    0 & \Delta_{12} & \Delta_{13}\\
    -\Delta_{12} & 0 & \Delta_{23} \\
    -\Delta_{13} & -\Delta_{23} & 0
    \end{pmatrix} ~,
\end{equation}
where the matrix $\Delta = O_H O_A^\text{T} - 1\!\!1$ captures the departure from the decoupling limit. We find
\begin{eqnarray}
\label{lambdamix_12}
    \Delta_{12} &=& \frac{v^2}{m_{A^\prime}^2} \frac{1}{t_\beta t_{\beta^\prime}} \left( 2 \lambda_3 + \lambda_4 - \lambda_5 - \lambda_6 +\lambda_7 - \lambda_8 - \lambda_9 \right) \equiv \frac{v^2}{m_{A^\prime}^2} \frac{1}{t_\beta t_{\beta^\prime}} \lambda_{12}~, \\ \label{lambdamix_13}
    \Delta_{13} &=& -\frac{v^2}{m_{A^\prime}^2} \frac{1}{t_{\beta^\prime}} \left( 2 \lambda_3 - \lambda_5 - \lambda_8 \right) \equiv -\frac{v^2}{m_{A^\prime}^2} \frac{1}{t_{\beta^\prime}} \lambda_{13}~, \\ \label{lambdamix_23}
    \Delta_{23} &=& -\frac{v^2}{m_A^2} \frac{1}{t_\beta} \left( 2 \lambda_3 - \lambda_6 - \lambda_9 \right) \equiv -\frac{v^2}{m_A^2} \frac{1}{t_\beta} \lambda_{23} ~.
\end{eqnarray}

\subsection{The Yukawa sector} \label{sec:yukawa}

We consider the following set of Yukawa matrices, with textures that are an extension of the textures previously explored in the context of so-called flavorful 2HDMs~\cite{Altmannshofer:2017uvs, Altmannshofer:2016zrn, Altmannshofer:2018bch}
\begin{subequations}
\begin{align} \label{eq:lambda_u}
    \lambda_{u_1}&\sim\frac{\sqrt{2}}{v_1}
    \begin{pmatrix}
    m_u & m_u & m_u\\
    m_u & m_u & m_u\\
    m_u & m_u & m_u
    \end{pmatrix}, 
    &\lambda_{u_2}&\sim\frac{\sqrt{2}}{v_2}
    \begin{pmatrix}
    0 & 0 & 0\\
    0 & m_c & m_c\\
    0 & m_c & m_c
    \end{pmatrix}, 
    &\lambda_{u_3}&\sim\frac{\sqrt{2}}{v_3}
    \begin{pmatrix}
    0 & 0 & 0\\
    0 & 0 & 0\\
    0 & 0 & m_t
    \end{pmatrix},
    \\  \label{eq:lambda_d}
    \lambda_{d_1}&\sim\frac{\sqrt{2}}{v_1}
    \begin{pmatrix}
    m_d & m_s\lambda & m_b\lambda^3\\
    m_d & m_d & m_d\\
    m_d & m_d & m_d
    \end{pmatrix}, 
    &\lambda_{d_2}&\sim\frac{\sqrt{2}}{v_2}
    \begin{pmatrix}
    0 & 0 & 0\\
    0 & m_s & m_b\lambda^2\\
    0 & m_s & m_s
    \end{pmatrix}, 
    &\lambda_{d_3}&\sim\frac{\sqrt{2}}{v_3}
    \begin{pmatrix}
    0 & 0 & 0\\
    0 & 0 & 0\\
    0 & 0 & m_b
    \end{pmatrix},
    \\  \label{eq:lambda_l}
    \lambda_{\ell_1}&\sim\frac{\sqrt{2}}{v_1}
    \begin{pmatrix}
    m_e & m_e & m_e\\
    m_e & m_e & m_e\\
    m_e & m_e & m_e
    \end{pmatrix}, 
    &\lambda_{\ell_2}&\sim\frac{\sqrt{2}}{v_2}
    \begin{pmatrix}
    0 & 0 & 0\\
    0 & m_\mu & m_\mu\\
    0 & m_\mu & m_\mu
    \end{pmatrix}, 
    &\lambda_{\ell_3}&\sim\frac{\sqrt{2}}{v_3}
    \begin{pmatrix}
    0 & 0 & 0\\
    0 & 0 & 0\\
    0 & 0 & m_\tau
    \end{pmatrix}.
\end{align}
\end{subequations}
With $\sim$ we indicate the order of the entries, keeping in mind that non-zero entries in the same matrix can differ by complex $\mathcal O(1)$ factors. 
We assume that all of the Yukawa couplings are rank-1 matrices, but otherwise do not contain any specific structure. Because of the assumed rank-1 structure, each Higgs doublet couples only to a single linear combination of the three generations and we dub such a scenario a ``generational'' 3HDM (or G3HDM). The rank-1 structure of the Yukawas can be realized in a straight-forward way by having the SM fermions mix with three separate generations of vector-like fermions that each couple to only one of the Higgs doublets. More details are given in appendix~\ref{app:rank1}.

An interesting feature of the generational 3HDM setup is that one may choose to generate part of the SM fermion mass hierarchies by adjusting the vevs of the three doublets relative to each other. This offers the option to either generate the large mass differences purely in the Yukawa sector, as seen in the SM, or to shift some of the burden over to the Higgs sector. In the limit of a large relative vev hierarchy, $v_1 \ll v_2 \ll v_3$, one may begin with a set of rank-1 and $\mathcal{O}(1)$ Yukawa couplings, and proceed to have the large parts of the quark and lepton mass hierarchies be generated in the Higgs sector alone. 

Without loss of generality, we have expressed the Yukawa couplings in a flavor basis in which the couplings of $\Phi_3$ are diagonal, emphasizing the role of $\Phi_3$ in providing the dominant component of the mass for the third generation of quarks and leptons. Note that the couplings of $\Phi_3$ preserve a $SU(2)^5$ flavor symmetry that acts on the first two generations. The couplings of $\Phi_2$ and $\Phi_1$ are in general misaligned in flavor space and both break the $SU(2)^5$ symmetry. However, we can use the remaining freedom in choosing a flavor basis to bring the couplings of $\Phi_2$ into the shown block-diagonal form. 

The non-minimal breaking of the SM flavor symmetries by the above set of Yukawa couplings is expected to give large contributions to flavor-changing neutral current (FCNC) processes, kaon and $D$ meson mixing in particular. The corresponding flavor constraints on the model will be analyzed in detail in section~\ref{sec:flavorprobe}. In order to soften the constraints one may entertain the possibility that some of the off-diagonal entries in $\lambda_{f_1}$ and $\lambda_{f_2}$ are suppressed compared to what is shown in~\eqref{eq:lambda_u} - \eqref{eq:lambda_l}. This could for example be achieved by spontaneously broken flavor symmetries.
However, we emphasize that not all off-diagonal entries in the Yukawa couplings can be arbitrarily suppressed. The entries in $\lambda_{d_1}$ and $\lambda_{d_2}$ of order $m_s \lambda$, $m_b \lambda^3$, and $m_b \lambda^2$, where $\lambda \simeq 0.23$ is the sine of the Cabibbo angle, are required to reproduce the CKM matrix, once one rotates in the fermion mass eigenstate basis. Note that it is an assumption that the CKM matrix originates from the diagonalization of the down-quark mass matrix. Other Yukawa textures can also be viable (see also footnote \ref{foot:up-sector} below.) 

In the fermion mass eigenstate basis, we define the following set of mass parameters which we will use below to write the couplings of the physical Higgs mass eigenstates to fermion mass eigenstates
\begin{equation}
    m^{f_1}_{ff^\prime}
    =
    \frac{v_1}{\sqrt{2}}
    \langle f_L|\lambda_{f_1}|f_{R}^\prime\rangle
    ~,~~~
    m^{f_2}_{ff^\prime}
    =
    \frac{v_2}{\sqrt{2}}
    \langle f_L|\lambda_{f_2}|f_{R}^\prime\rangle
    ~,~~~
    m^{f_3}_{ff^\prime}
    =
    \frac{v_3}{\sqrt{2}}
    \langle f_L|\lambda_{f_3}|f_{R}^\prime\rangle
    ~.
\end{equation}
These parameters obey the relationship
\begin{equation}
    m^{f_3}_{ff^\prime}
    +
    m^{f_2}_{ff^\prime}
    +
    m^{f_1}_{ff^\prime}    
    =
    m_{f}\delta_{ff^\prime}
    ~,
\end{equation}
with $m_f$ (without superscripts) representing the physical masses of the quarks or leptons.

We obtain the following set of explicit mass parameters after expanding each entry to leading order in ratios of first to second and second to third generation masses
\begin{equation} \label{eq:mu1and2}
    \frac{m^{u_1}_{qq^\prime}}{m_u} \simeq \begin{pmatrix}
       1 & x_{uc} & x_{ut} \\
       x_{cu} & x_{cu} x_{uc} & x_{cu} x_{ut} \\
       x_{tu} & x_{tu} x_{uc} & x_{tu} x_{ut}
    \end{pmatrix} , ~~~
    \frac{m^{u_2}_{qq^\prime}}{m_c} \simeq \begin{pmatrix}
       \frac{m_u^2}{m_c^2} x_{uc} x_{cu} & - \frac{m_u}{m_c} x_{uc} & -\frac{m_u}{m_c} x_{uc} y_{ct} \\
       -\frac{m_u}{m_c} x_{cu} & 1 & y_{ct} \\
       -\frac{m_u}{m_c} y_{tc} x_{cu} & y_{tc} & y_{tc} y_{ct}
    \end{pmatrix} ~,
\end{equation}
\begin{equation} \label{eq:mu3}
    \frac{m^{u_3}_{qq^\prime}}{m_t} \simeq \begin{pmatrix}
       \frac{m_u^2}{m_t^2} (x_{ut} - x_{uc} y_{ct}) (x_{tu} - y_{tc} x_{cu}) & \frac{m_u m_c}{m_t^2} (x_{ut} - x_{uc} y_{ct}) y_{tc} & -\frac{m_u}{m_t} (x_{ut} - x_{uc} y_{ct}) \\
       \frac{m_u m_c}{m_t^2} y_{ct} (x_{tu} - y_{tc} x_{cu}) & \frac{m_c^2}{m_t^2} y_{ct} y_{tc} & - \frac{m_c}{m_t} y_{ct} \\
       -\frac{m_u}{m_t} (x_{tu} - y_{tc} x_{cu}) & - \frac{m_c}{m_t} y_{tc} & 1
    \end{pmatrix} ~,
\end{equation}
\begin{equation}\label{eq:md1and2}
    \frac{m^{d_1}_{qq^\prime}}{m_d} \simeq \begin{pmatrix}
       1 & \frac{m_s}{m_d} V_{ud}^* V_{us} & \frac{m_b}{m_d} V_{ud}^* V_{ub} \\
       x_{sd} & x_{sd} \frac{m_s}{m_d} V_{ud}^* V_{us} & x_{sd} \frac{m_b}{m_d} V_{ud}^* V_{ub}\\
       x_{bd} & x_{bd} \frac{m_s}{m_d} V_{ud}^* V_{us} & x_{bd} \frac{m_b}{m_d} V_{ud}^* V_{ub}
    \end{pmatrix} , ~~~
    \frac{m^{d_2}_{qq^\prime}}{m_s} \simeq \begin{pmatrix}
       -\frac{m_d}{m_s} V_{cd}^* V_{cs} x_{sd} & V_{cd}^* V_{cs} & \frac{m_b}{m_s} V_{cd}^* V_{cb} \\
       -\frac{m_d}{m_s} x_{sd} & 1 & \frac{m_b}{m_s} V_{cs}^* V_{cb} \\
       -\frac{m_d}{m_s} y_{bs} x_{sd} & y_{bs} & y_{bs} \frac{m_b}{m_s} V_{cs}^* V_{cb}
    \end{pmatrix} ~,
\end{equation}
\begin{equation}\label{eq:md3}
    \frac{m^{d_3}_{qq^\prime}}{m_b} \simeq \begin{pmatrix}
       -\frac{m_d}{m_b} V_{td}^*V_{tb} (x_{bd} - y_{bs} x_{sd}) & - \frac{m_s}{m_b} V_{td}^* V_{tb} y_{bs} & V_{td}^*  V_{tb} \\
       -\frac{m_d}{m_b} V_{ts}^* V_{tb} (x_{bd} - y_{bs} x_{sd}) & -\frac{m_s}{m_b} V_{ts}^* V_{tb} y_{bs} & V_{ts}^* V_{tb} \\
       -\frac{m_d}{m_b} (x_{bd} - y_{bs} x_{sd}) & - \frac{m_s}{m_b} y_{bs} & 1
    \end{pmatrix} ~.
\end{equation}
The lepton mass parameters are completely analogous to the ones in the up quark sector. 
In the above expressions, the $x_{ij}$, $y_{ij}$ are free, in general complex, $\mathcal O(1)$ parameters. These parameterize new sources of flavor and CP violation. Note that not all entries of the mass matrices are independent. This is due to two reasons: first, they need to reproduce the quark and lepton masses as well as the CKM matrix; second, we have assumed that they originate from rank-1 Yukawa couplings.

It is in principle possible to set all the $x_{ij}$, $y_{ij}$ to zero. In that case, one obtains a ``generation specific'' 3HDM in which the Yukawa couplings are aligned such that $\Phi_1$, $\Phi_2$, and $\Phi_3$ couple to a good approximation only to the first, second, and third generation, respectively. The alignment can be exact in the up-quark sector and the lepton sector. However, because in the described setup the CKM matrix originates from the down Yukawa couplings, the alignment cannot be perfect in the down sector. The mass matrices $m^{d_3}_{qq^\prime}$, $m^{d_2}_{qq^\prime}$, $m^{d_1}_{qq^\prime}$ contain necessarily also flavor changing entries\footnote{Alternatively, the CKM matrix could be generated in the up-sector. In that case one could have an exact alignment in the down quark sector, but the up quark mass matrices $m^{u_3}_{qq^\prime}$, $m^{u_2}_{qq^\prime}$, $m^{u_1}_{qq^\prime}$ would contain off-diagonal terms. In the most generic case, the CKM matrix would originate partly from the down-sector and partly from the up-sector. \label{foot:up-sector}}. 

\subsection{Couplings of the physical Higgs bosons}

After bringing both fermions and scalars into the mass eigenstate basis, the couplings of the physical Higgs bosons of the 3HDM to the SM fermions may be parameterized by
\begin{multline}
    -\mathcal{L}_{Y} \supset
    \sum\limits_{f=d,\ell} \sum\limits_{i,j} 
    (\overline{f}_i P_R f_j)
    \Big(
    h \kappa^{h}_{f_i f_j} 
    +
    H \kappa^{H}_{f_i f_j} 
    +
    H^\prime \kappa^{H^\prime}_{f_i f_j} 
    +
    i A \kappa^{A}_{f_i f_j} 
    +
    i A^\prime \kappa^{A^\prime}_{f_i f_j} 
    \Big) \frac{m_{f_j}}{v} + \text{h.c.}
    \\
    + \sum\limits_{i,j} 
    (\overline{u}_i P_R u_j) 
    \Big(
    h \kappa^{h}_{u_i u_j} 
    +
    H \kappa^{H}_{u_i u_j} 
    +
    H^\prime \kappa^{H^\prime}_{u_i u_j} 
    -
    i A \kappa^{A}_{u_i u_j} 
    -
    i A^\prime \kappa^{A^\prime}_{u_i u_j} 
    \Big) \frac{m_{u_j}}{v} + \text{h.c.}
    \\
    + \sqrt{2} \sum\limits_{i,j}  (\overline{\nu}_i P_R \ell_j)
    \Big(H^{+} \kappa^{\pm}_{\ell_i \ell_j} 
    +
    H^{+\,\prime} \kappa^{\pm\,\prime}_{\ell_i \ell_j}
    \Big) \frac{m_{\ell_j}}{v} + \text{h.c.}
    \\
    + \sqrt{2} \sum\limits_{i,j,k}  (\overline{u}_i P_R d_j)  V_{u_i d_k}
    \Big(H^{+} \kappa^{\pm}_{d_k d_j} 
    +
    H^{+\,\prime} \kappa^{\pm\,\prime}_{d_k d_j}
    \Big) \frac{m_{d_j}}{v} + \text{h.c.}
    \\
    - \sqrt{2} \sum\limits_{i,j,k}  (\overline{d}_i P_R u_j)  V^*_{u_k d_i}
    \Big(H^{-} \kappa^{\pm}_{u_k u_j} 
    +
    H^{-\,\prime} \kappa^{\pm\,\prime}_{u_k u_j}
    \Big)\frac{m_{u_j}}{v} + \text{h.c.} ~.
\end{multline}
We introduced the $\kappa$ modifiers that parameterize the deviation from the related SM couplings. In the SM, $\kappa^h_{f_i f_i}=1$ for the diagonal Higgs couplings, whereas $\kappa^h_{f_i f_j}=0$ for the off-diagonal couplings.
It is straight-forward to express the $\kappa$ factors in terms of mass parameters in~\eqref{eq:mu1and2} - \eqref{eq:md3} and the mixing angles in the Higgs sector defined in~\eqref{eq:OA}, \eqref{eq:Opm}, and \eqref{eq:OH},
\begin{equation}
    \begin{pmatrix}
    \kappa^{A^\prime}_{ff^\prime} \frac{m_{f^\prime}}{v} \\ \kappa^{A}_{ff^\prime} \frac{m_{f^\prime}}{v} \\ \delta_{ff^\prime} \frac{m_{f^\prime}}{v}
    \end{pmatrix} = O_{A}
    \begin{pmatrix}
    \frac{m^{f_1}_{ff^\prime}}{v_1} \\ \frac{m^{f_2}_{ff^\prime}}{v_2} \\ \frac{m^{f_3}_{ff^\prime}}{v_3}
    \end{pmatrix}
    ~, ~~~
    \begin{pmatrix}
    \kappa^{H^\prime}_{ff^\prime} \frac{m_{f^\prime}}{v}  \\ \kappa^{H}_{ff^\prime} \frac{m_{f^\prime}}{v}  \\ \kappa^{h}_{ff^\prime} \frac{m_{f^\prime}}{v} 
    \end{pmatrix} = O_{H}
    \begin{pmatrix}
    \frac{m^{f_1}_{ff^\prime}}{v_1} \\ \frac{m^{f_2}_{ff^\prime}}{v_2} \\ \frac{m^{f_3}_{ff^\prime}}{v_3}
    \end{pmatrix}
    ~, ~~~
    \begin{pmatrix}
    \kappa^{\pm\,\prime}_{ff^\prime} \frac{m_{f^\prime}}{v}  \\ \kappa^{\pm}_{ff^\prime} \frac{m_{f^\prime}}{v}  \\ \delta_{ff^\prime} \frac{m_{f^\prime}}{v} 
    \end{pmatrix} = O_{\pm}
    \begin{pmatrix}
    \frac{m^{f_1}_{ff^\prime}}{v_1} \\ \frac{m^{f_2}_{ff^\prime}}{v_2} \\ \frac{m^{f_3}_{ff^\prime}}{v_3}
    \end{pmatrix}
    ~.
\end{equation}
As the resulting explicit expressions are very lengthy, we refrain from showing them here. Instead, we give approximate expressions that are valid in the combined limit $v^2 \ll m_A^2, m_{A^\prime}^2$ and $1 \ll \tan\beta \ll \tan\beta^\prime$. We find for the CP-odd Higgs couplings
\begin{eqnarray}
    \kappa^{A^\prime}_{f_i f_j} m_{f_j} &\simeq& t_{\beta^\prime} m^{f_1}_{f_i f_j} - t_{\beta} t_\gamma m^{f_2}_{f_i f_j} - \frac{1}{t_{\beta^\prime}} m^{f_3}_{f_i f_j}~, \\
    \kappa^{A}_{f_i f_j} m_{f_j} &\simeq& t_{\beta} m^{f_2}_{f_i f_j} + t_{\beta^\prime} t_\gamma m^{f_1}_{f_i f_j} - \frac{1}{t_{\beta}} m^{f_3}_{f_i f_j}~.
\end{eqnarray}
The $\kappa$ factors of the CP even Higgs bosons are tightly related to the ones of the CP odd Higgs bosons. For all types of fermions one has
\begin{eqnarray}
    \kappa^{H^\prime}_{f_i f_j} &\simeq& \kappa^{A^\prime}_{f_i f_j} - \frac{v^2}{m_{A^\prime}^2} \frac{\lambda_{13}}{t_{\beta^\prime}} \delta_{ij} + \frac{v^2}{m_{A^\prime}^2}  \frac{\lambda_{12}}{t_{\beta^\prime} t_\beta}\kappa^{A}_{f_i f_j}  ~, \\
    \kappa^{H}_{f_i f_j} &\simeq& \kappa^{A}_{f_i f_j} - \frac{v^2}{m_A^2} \frac{\lambda_{23}}{t_\beta} \delta_{ij} - \frac{v^2}{m_{A^\prime}^2} \frac{\lambda_{12}}{t_{\beta^\prime} t_\beta} \kappa^{A^\prime}_{f_i f_j} ~, \\
    \kappa^{h}_{f_i f_j} &\simeq& \delta_{ij} + \frac{v^2}{m_A^2} \frac{\lambda_{23}}{t_\beta} \kappa^A_{f_i f_j} +\frac{v^2}{m_{A^\prime}^2} \frac{\lambda_{13}}{t_{\beta^\prime}} \kappa^{A^\prime}_{f_i f_j} ~,
\end{eqnarray}
with the parameters $\lambda_{12}$, $\lambda_{13}$, $\lambda_{23}$ defined in eqs.~\eqref{lambdamix_12} - \eqref{lambdamix_23}. 

Finally, for the charged Higgs bosons we find
\begin{equation}
    \kappa^{H^{\pm \, \prime}}_{f_i f_j} \simeq \kappa^{A^\prime}_{f_i f_j} - \frac{t_\gamma v^2}{2 m_{A^\prime}^2} (\lambda_8 - \lambda_9 ) \kappa^{A}_{f_i f_j} ~,~~~
    \kappa^{H^\pm}_{f_i f_j} \simeq \kappa^{A}_{f_i f_j} + \frac{t_\gamma v^2}{2 m_{A^\prime}^2} (\lambda_8 - \lambda_9 ) \kappa^{A^\prime}_{f_i f_j} ~.
\end{equation}

\section{Low Energy Flavor Probes}
\label{sec:flavorprobe}

As seen in the previous section, the Higgs bosons of the considered setup in general all have flavor-changing couplings. One thus expects that flavor-changing neutral current (FCNC) processes can be used to constrain the model. This is particularly the case for neutral meson mixing and rare leptonic decays of neutral mesons, which all receive tree-level contributions from neutral Higgs boson exchange.
As detailed below, we indeed find that observables related to the mentioned FCNC processes provide often stringent constraints on our Higgs parameter space as many of the observables have been measured to a high accuracy and there exist robust SM predictions for most cases considered. 

\subsection{Meson oscillations} \label{sec:mixing}

The frequencies of neutral kaon and $B$ meson oscillations are measured with impressive accuracy~\cite{LHCb:2016gsk, LHCb:2021moh, ParticleDataGroup:2024cfk, HeavyFlavorAveragingGroup:2022wzx}. 
Also the parameter $\epsilon_K$ that measures indirect CP violation in kaon oscillations is known experimentally with high precision~\cite{ParticleDataGroup:2024cfk}, as are the $B^0$ and $B_s$ mixing phases~\cite{HeavyFlavorAveragingGroup:2022wzx, LHCb:2023zcp, LHCb:2023sim}. The experimental status is summarized in the left column of Table~\ref{tab:mesonexp}.

\setlength{\tabcolsep}{12pt}
\renewcommand{\arraystretch}{1.3}
\begin{table}[tb]
    \begin{center}
    \begin{tabular}{ccc}
    \hline \hline
    observable &
    experiment &
    SM prediction (from tree level CKM) \\
    \hline \hline
    $\Delta M_{K}$ & $\left(3.484 \pm 0.006 \right)\times10^{-15}~\text{GeV}$~\cite{ParticleDataGroup:2024cfk} & $\left(3.1 \pm 1.2\right)\times10^{-15}~\text{GeV}$~\cite{Brod:2011ty} \\
    $\Delta M_{B^0}$ & $(0.5069\pm0.0019)~\text{ps}^{-1}$~\cite{HeavyFlavorAveragingGroup:2022wzx} & $(0.481\pm0.040)~\text{ps}^{-1}$ \\
    $\Delta M_{B_s}$ & $(17.765\pm0.006)~\text{ps}^{-1}$~\cite{HeavyFlavorAveragingGroup:2022wzx} & $(16.62\pm1.14)~\text{ps}^{-1}$  \\
    \hline
    $\epsilon_{K}$ & $\left(2.228\pm{0.011}\right)\times10^{-3}$~\cite{ParticleDataGroup:2024cfk} & $\left(2.10\pm{0.20}\right)\times10^{-3}$ \\
    $\phi_d$ & $45.2^\circ \pm 0.9^\circ$~\cite{HeavyFlavorAveragingGroup:2022wzx, LHCb:2023zcp} & $47.5^\circ \pm 3.1^\circ$ \\
    $\phi_s$ &  $-2.29^\circ \pm 0.92^\circ$~\cite{HeavyFlavorAveragingGroup:2022wzx, LHCb:2023sim} & $-2.18^\circ\pm0.14^\circ$ \\
    \hline \hline
    \end{tabular}
    \caption{Experimental measurements and SM predictions for observables in neutral meson oscillations. Note that the SM prediction for $\Delta M_K$ corresponds to the short-distance contribution only. To account for possible sizable long-distance effects (see~\cite{Bai:2014cva,Wang:2022lfq} for first attempts to calculate those on the lattice), we use instead $\Delta M_K^\text{SM} = \Delta M_K^\text{exp} (1 \pm 0.5)$ in our numerical analysis. The SM predictions without reference are based on our own numerical evaluation. See the text for details.}
    \label{tab:mesonexp}
    \end{center}
\end{table}

The SM predictions we use are collected in the right column of Table~\ref{tab:mesonexp}. 
For the neutral kaon oscillation frequency $\Delta M_K$, we quote the short-distance contribution~\cite{Brod:2011ty}. Keeping in mind that there are also long-distance contributions that are poorly controlled so far~\cite{Bai:2014cva, Wang:2022lfq}, we use in our numerical analysis $\Delta M_K^\text{SM} = \Delta M_K^\text{exp} (1 \pm 0.5)$. We obtain the SM predictions of the remaining observables following~\cite{Altmannshofer:2021uub}. Among the most relevant input parameters are CKM matrix elements that we determine from the PDG values~\cite{ParticleDataGroup:2024cfk}
\begin{equation} \label{eq:CKM_input}
    |V_{cb}| = (41.1 \pm 1.2)\times 10^{-3}~,~~~ |V_{ub}| = (3.82 \pm 0.20)\times 10^{-3}~,~~~ \gamma = 65.7^\circ \pm 3.0^\circ~.
\end{equation}
The values for $|V_{cb}|$ and $|V_{ub}|$ are conservative averages of determinations using inclusive and exclusive tree level $B$ decays.
For the sine of the Cabibbo angle, we use $\lambda \simeq 0.225$~\cite{ParticleDataGroup:2024cfk}, neglecting its tiny uncertainty. The hadronic matrix elements needed for the SM predictions of $\Delta M_d$ and $\Delta M_s$ are taken from~\cite{Dowdall:2019bea}. 
Overall, there is very good agreement between the measurements and the corresponding SM predictions. In most cases, the size of possible new physics contributions is limited by the precision of the SM predictions.

The relevant observables in the neutral $B$ meson systems are the mass differences $\Delta M_q$ and mixing phases $\phi_q$ that can be calculated from the new physics contributions to the mixing amplitudes, which we denote by $M_{12}^\text{NP}$
\begin{equation}
    \Delta M_q = \Delta M_{q}^{\text{SM}} \times \left|1+\frac{M_{12}^{\text{NP}}}{M_{12}^{\text{SM}}} \right|
    ~,~~~
    \phi_q = \phi_q^{\text{SM}} + \text{Arg}\Big( 1+\frac{M_{12}^{\text{NP}}}{M_{12}^{\text{SM}}} \Big) ~.
\end{equation}
Similarly, the relevant observables for kaon mixing are the mass difference $\Delta M_K$ and the CP violating parameter $\epsilon_K$, and they can be expressed as
\begin{equation}
    \Delta M_K = \Delta M_{K}^{\text{SM}} + 2 \, \text{Re} \left(M_{12}^{\text{NP}}\right)
    ~,~~~
    \epsilon_K = \epsilon_K^{\text{SM}} + \sin(\phi_e) \frac{\text{Im} \left(M_{12}^{\text{NP}}\right)}{\sqrt{2}\Delta M_K}
    ~,
\end{equation}
where $\phi_e = (43.52 \pm 0.05)^\circ$~\cite{ParticleDataGroup:2024cfk}.

In the considered 3HDM setup, the dominant new physics contributions arise from the tree-level exchange of the neutral Higgs bosons. The expressions for $B^0$ and $B_s$ mixing are an extension of the flavorful 2HDM expressions~\cite{Altmannshofer:2018bch}, and we find

\begin{multline} \label{eq:M12_Bmixing}
    \frac{M_{12}^\text{NP}}{M_{12}^\text{SM}} = \frac{16\pi^2}{g^2} \frac{1}{S_0} \Bigg[ 2 X_4 \sum_{i=h,H,H^\prime,A,A^\prime} \frac{m_q}{m_b} \frac{\big(\kappa^{i\,*}_{bq}\big)\big(\kappa^{i}_{qb}\big)}{(V_{tq}^* V_{tb})^2}\frac{m_{B_q}^2}{m_{i}^2} \\
    + \Big( X_2 + X_3 \Big) \bigg( \sum_{i=h,H,H^\prime} - \sum_{i=A,A^\prime} \bigg)  \left(\frac{\big(\kappa^{i}_{qb}\big)^2}{(V_{tq}^* V_{tb})^2} + \frac{m_q^2}{m_b^2} \frac{\big(\kappa^{i\,*}_{bq}\big)^2}{(V_{tq}^* V_{tb})^2} \right) \frac{m_{B_q}^2}{m_i^2} \Bigg] ~,
\end{multline}
where $S_0 \simeq 2.31$ is a SM loop function. 
The parameters $X_2$, $X_3$, and $X_4$ encapsulate 1-loop QCD renormalization group running and ratios of hadronic matrix elements. 
Setting the renormalization scale to $\mu = 1$\,TeV and using hadronic matrix elements from~\cite{FermilabLattice:2016ipl}, we find 
\begin{eqnarray}
  &&  X_2^d \simeq -0.42~,\qquad X_3^d \simeq -0.0056 ~,\qquad  X_4^d \simeq 1.14 ~, \\
  &&  X_2^s \simeq -0.43~,\qquad X_3^s \simeq -0.0055 ~,\qquad  X_4^s \simeq 1.07 ~,
\end{eqnarray}
for $B^0$ mixing and for $B_s$ mixing, respectively. Natural choices for the renormalization scale are the masses of the neutral Higgs bosons. The values of the $X_i$ depend logarithmically on the scale and stay within $\pm 10\%$ when varying it between $\mu = 100$\,GeV and $\mu = 10$\,TeV. 
Explicit expressions for the $X_i$ factors are provided in appendix~\ref{appendix:mix}.

The final ingredient to evaluate eq.~\eqref{eq:M12_Bmixing} are the quark mass ratios $m_q/m_b$. The ratios are to a very good approximation RGE invariant, and we use~\cite{ParticleDataGroup:2024cfk, Chetyrkin:2000yt, Herren:2017osy}
\begin{equation}
    m_d/m_s \simeq 5.0 \times 10^{-2} ~,\qquad m_d/m_b \simeq 9.4 \times 10^{-4} ~,\qquad m_s/m_b \simeq 1.9 \times 10^{-2} ~. 
\end{equation}

In the phenomenologically interesting limit $v \ll m_A, m_{A^\prime}$ there is an approximate cancellation in the terms in the second line of eq.~\eqref{eq:M12_Bmixing}. In this case, using the expressions for the $\kappa$ factors from section~\ref{sec:G3HDM}, we find
\begin{eqnarray}
\label{BdmixM12}
 B^0 ~\text{mixing}: \quad &&  \frac{M_{12}^\text{NP}}{M_{12}^\text{SM}} \simeq \frac{16\pi^2}{g^2} \frac{1}{S_0} 4 X^d_4 \frac{m_d}{m_b} \left( t_{\beta^\prime}^2 \frac{m_{B^0}^2}{m_{A^\prime}^2} \frac{x_{bd}^* V_{ud}^* V_{ub}}{(V_{td}^* V_{tb})^2} - t_\beta^2 \frac{m_{B^0}^2}{m_A^2} \frac{y_{bs}^* x_{sd}^* V_{cd}^* V_{cb}}{(V_{td}^* V_{tb})^2} \right) ~, \\
 \label{BsmixM12}
 B_s ~\text{mixing}: \quad &&
 \frac{M_{12}^\text{NP}}{M_{12}^\text{SM}} \simeq  \frac{16\pi^2}{g^2} \frac{1}{S_0} 4 X^s_4 \frac{m_s}{m_b} \left( t_{\beta^\prime}^2 \frac{m_{B_s}^2}{m_{A^\prime}^2} \frac{x_{bd}^* x_{sd} V_{us}^* V_{ub}}{(V_{ts}^* V_{tb})^2} - t_\beta^2 \frac{m_{B_s}^2}{m_A^2} \frac{y_{bs}^*}{V_{ts}^* V_{tb}} \right) ~.
\end{eqnarray}

Based on the above expressions and using the values collected in Table~\ref{tab:mesonexp}, we can derive simple analytical bounds on $m_A$ and $m_{A^\prime}$. Assuming that there are no accidental cancellations, setting the absolute values of the $x_{ij}$ and $y_{ij}$ parameters to 1, and marginalizing over their phases, we find
\begin{align}
m_A \gtrsim \tan\beta \times 492 ~\text{GeV} ~&,~~~ m_{A^\prime} \gtrsim \tan\beta^\prime \times 72 ~\text{GeV} ~,~~~ \text{from} ~B_s~ \text{mixing}~, \\
m_A \gtrsim \tan\beta \times 147 ~\text{GeV} ~&,~~~ m_{A^\prime} \gtrsim \tan\beta^\prime \times 93 ~\text{GeV} ~,~~~ \text{from} ~B^0~ \text{mixing}~.
\end{align}

For moderately large values of $\tan\beta \sim 5$ and $\tan\beta^\prime \sim 25$, we find that the masses of the additional Higgs bosons get pushed into the multi-TeV range.

In the scenario in which all $x_{ij} = y_{ij} =0$, the expressions in eqs.~\eqref{BdmixM12} and \eqref{BsmixM12} vanish, and one needs to expand one order higher in $v^2/m_{A^{(\prime)}}^2$ to find the leading contribution. An analogous behavior was observed in the context of flavorful 2HDMs~\cite{Altmannshofer:2018bch} and is a well-known phenomenon in 2HDMs in general~\cite{Gorbahn:2009pp}; see also~\cite{Crivellin:2013wna}. Making use of the results in section~\ref{sec:beyond_decoupling_limit}, we find
\begin{eqnarray}
\label{B0mixing}
B^0 ~\text{mixing}: \quad &&      
    \frac{M_{12}^\text{NP}}{M_{12}^\text{SM}} \simeq \frac{16\pi^2}{g^2} \frac{1}{S_0} \Big(X_2^d+X_3^d\Big) \Bigg[ \frac{m_{B^0}^2 v^2}{m_A^4} \frac{(V_{cd}^* V_{cb})^2}{(V_{td}^* V_{tb})^2} \left( \frac{\lambda_{23}^2}{2\lambda_3} - 2 \lambda_H \right) \qquad\qquad\qquad \nonumber \\
    && \qquad\qquad\qquad\qquad\qquad\quad + \frac{m_{B^0}^2 v^2}{m_{A^\prime}^4} \frac{(V_{ud}^* V_{ub})^2}{(V_{td}^* V_{tb})^2} \left( \frac{\lambda_{13}^2}{2\lambda_3} - 2 \lambda_{H^\prime} \right) \nonumber \\
    && \qquad\qquad\qquad\qquad\qquad\quad + \frac{2 m_{B^0}^2 v^2}{m_{A^\prime}^2 m_A^2} \frac{(V_{ud}^* V_{ub})(V_{cd}^* V_{cb})}{(V_{td}^* V_{tb})^2} \left( \frac{\lambda_{13} \lambda_{23}}{2\lambda_3} - \lambda_{12} \right) \Bigg]~, \\[12pt]
\label{Bsmixing}
B_s ~\text{mixing}: \quad &&           
    \frac{M_{12}^\text{NP}}{M_{12}^\text{SM}} \simeq  \frac{16\pi^2}{g^2} \frac{1}{S_0} \Big(X_2^s+X_3^s\Big) \frac{m_{B_s}^2 v^2}{m_A^4} \left( \frac{\lambda_{23}^2}{2\lambda_3} - 2 \lambda_H \right) ~,
\end{eqnarray}
Interestingly, in this case, the expressions are independent of $\tan\beta$ and $\tan\beta^\prime$, and one can directly obtain bounds on the masses $m_A$ and $m_{A^\prime}$. Assuming the absence of accidental cancellations and assuming that the relevant combinations of quartic coupling are 1, we find
\begin{eqnarray} \label{eq:Bs_mixing_constraint}
m_A \gtrsim  206 ~\text{GeV} ~,~~~ \phantom{m_{A^\prime} \gtrsim  ..... ~\text{GeV} ~,} && ~~~ \text{from} ~B_s~ \text{mixing}~, \\ \label{eq:B0_mixing_constraint}
m_A \gtrsim  215 ~\text{GeV} ~,~~~ m_{A^\prime} \gtrsim  102 ~\text{GeV} ~, && ~~~ \text{from} ~B^0~ \text{mixing}~.
\end{eqnarray}

We do not obtain a meaningful bound on $m_{A^\prime}$ from $B_s$ mixing. The bounds that can be obtained are rather weak, of the order of the electroweak scale. Given our approximations, we expect $\mathcal O(1)$ uncertainties on those bounds.  

Moving on to kaon oscillations, the new physics contributions to the mixing amplitude can be generically written as
\begin{multline}
        M_{12}^\text{NP} = m_K \frac{f_K^2}{v^2} \Bigg[ \frac{1}{4} B_4^K \eta_4 \sum_{i=h,H,H^\prime,A,A^\prime} \frac{m_d}{m_s} \big(\kappa^{i\,*}_{sd}\big)\big(\kappa^{i}_{ds}\big) \frac{m_K^2}{m_{i}^2} \\
    - \bigg( \frac{5}{48} B_2^K \eta_2 - \frac{1}{48} B_3^K \eta_3 \bigg) \bigg( \sum_{i=h,H,H^\prime} - \sum_{i=A,A^\prime} \bigg)  \left(\big(\kappa^{i}_{ds}\big)^2 + \frac{m_d^2}{m_s^2} \big(\kappa^{i\,*}_{sd}\big)^2 \right) \frac{m_K^2}{m_i^2} \Bigg] ~.
\end{multline}
For the kaon decay constant and the hadronic bag parameters we use $f_K \simeq 155.7$~MeV~\cite{FlavourLatticeAveragingGroupFLAG:2021npn} and $B^K_2 \simeq 0.46$, $B^K_3 \simeq 0.79$, $B^K_4 \simeq 0.78$~\cite{Carrasco:2015pra}. 
The $\eta$ factors correspond to corrections from the 1-loop QCD renormalization group running from the high new physics scale to the low scale at which the kaon matrix elements are evaluated. For a new physics scale of $\mu = 1$~TeV we find
\begin{equation} \label{eq:etas}
    \eta_2 \simeq 0.64 ~,\qquad \eta_3 \simeq -0.032 ~,\qquad \eta_4 = 1 ~.
\end{equation}
Explicit expressions for the $\eta$ factors are given in appendix~\ref{appendix:mix}. 

In the decoupling limit $v \ll m_A, m_{A^\prime}$, the expression simplifies considerably
\begin{equation}
\label{Kmixing}
        M_{12}^\text{NP} \simeq m_K \frac{f_K^2}{v^2} \frac{1}{2} B_4^K \eta_4 \frac{m_d}{m_s} x_{sd}^* V_{ud}^* V_{us} \left( t_{\beta^\prime}^2 \frac{m_K^2}{m_{A^\prime}^2} + t_\beta^2 \frac{m_K^2}{m_A^2} \right) ~.
\end{equation}
Setting the absolute value of the parameter $x_{sd}$ to 1 and marginalizing over its phase, we find the following bounds on the Higgs masses 
\begin{eqnarray}
    m_A \gtrsim \tan\beta \times 8.8 ~\text{TeV} ~,~~~ m_{A^\prime} \gtrsim \tan\beta^\prime \times 8.8 ~\text{TeV} ~, && ~~~ \text{from} ~K~ \text{mixing}~,
\end{eqnarray}
In the generic case, we see that the additional Higgs bosons are far outside the reach of the LHC, even for moderate values of $\tan\beta$ and $\tan\beta^\prime$.

If we profile over the phase of $x_{sd}$ instead of marginalizing over it, the situation is qualitatively different, as the strong constraint from $\epsilon_K$ is avoided and only the much weaker constraint from $\Delta M_K$ applies. In this approach, we find
\begin{eqnarray}
  m_A \gtrsim  \tan\beta \times 360~\text{GeV} ~,~~~ m_{A^\prime} \gtrsim  \tan\beta^\prime \times 360~\text{GeV} ~, && ~~~ \text{from} ~K~ \text{mixing}~.
\end{eqnarray}
If the $x_{sd}$ parameter is set to 0, the leading contribution to the kaon mixing amplitude is given by
\begin{multline}
        M_{12}^\text{NP} \simeq - m_K f_K^2 \bigg( \frac{5}{48} B_2^K \eta_2 - \frac{1}{48} B_3^K \eta_3 \bigg)(V_{ud}^* V_{us})^2 \Bigg[ \frac{m_K^2}{m_A^4} \left( \frac{\lambda_{23}^2}{2\lambda_3} - 2 \lambda_H \right) \\ + \frac{m_K^2}{m_{A^\prime}^4}  \left( \frac{\lambda_{13}^2}{2\lambda_3} - 2 \lambda_{H^\prime} \right)
        - \frac{2 m_K^2}{m_{A^\prime}^2 m_A^2} \left( \frac{\lambda_{13} \lambda_{23}}{2\lambda_3} - \lambda_{12} \right) \Bigg] ~,
\end{multline}
If the quartic couplings are assumed to be $1$ and barring accidental cancellations, we find
\begin{eqnarray}
  m_A \gtrsim  230~\text{GeV} ~,~~~ m_{A^\prime} \gtrsim  230~\text{GeV} ~, && ~~~ \text{from} ~K~ \text{mixing}~.
\end{eqnarray}
These constraints are fairly weak and in the same ballpark as the numbers we found from $B^0$ and $B_s$ mixing when the $x_{ij}$ and $y_{ij}$ were switched off.

For completeness, we also discuss the constraints from neutral $D$ meson mixing. The expressions for the mixing amplitude are very similar to the case of kaon mixing, with the obvious replacements of couplings, masses, and hadronic parameters
\begin{multline}
        M_{12}^\text{NP} = m_D \frac{f_D^2}{v^2} \Bigg[ \frac{1}{4} B_4^D \eta_4 \sum_{i=h,H,H^\prime,A,A^\prime} \frac{m_u}{m_c} \big(\kappa^{i\,*}_{cu}\big)\big(\kappa^{i}_{uc}\big) \frac{m_D^2}{m_{i}^2} \\
    - \bigg( \frac{5}{48} B_2^D \eta_2 - \frac{1}{48} B_3^D \eta_3 \bigg) \bigg( \sum_{i=h,H,H^\prime} - \sum_{i=A,A^\prime} \bigg)  \left(\big(\kappa^{i}_{uc}\big)^2 + \frac{m_u^2}{m_c^2} \big(\kappa^{i\,*}_{cu}\big)^2 \right) \frac{m_D^2}{m_i^2} \Bigg] ~.
\end{multline}
The $\eta$ parameters have already been introduced in eq.~\eqref{eq:etas}. The $D$ meson decay constant, and the bag parameters are $f_D \simeq 211.6$~MeV~\cite{FlavourLatticeAveragingGroupFLAG:2021npn} and $B^D_2 \simeq 0.65$, $B^D_3 \simeq 0.96$, $B^D_4 \simeq 0.91$~\cite{Carrasco:2015pra}. For the ratio of up to charm quark mass, we use~\cite{ParticleDataGroup:2024cfk, Chetyrkin:2000yt, Herren:2017osy}
\begin{equation}
    m_u/m_c \simeq 2.0 \times 10^{-3}~.
\end{equation}
In the decoupling limit, the expression for the mixing amplitude simplifies to
\begin{equation}
        M_{12}^\text{NP} \simeq m_D \frac{f_D^2}{v^2} \frac{1}{2} B_4^D \eta_4 \frac{m_u^2}{m_c^2} x_{cu}^* x_{uc} \left( t_{\beta^\prime}^2 \frac{m_D^2}{m_{A^\prime}^2} + t_\beta^2 \frac{m_D^2}{m_A^2} \right) ~.
\end{equation}
On the experimental side, neutral $D$ meson mixing is firmly established~\cite{HeavyFlavorAveragingGroup:2022wzx}. Assuming no direct CP violation in doubly Cabibbo suppressed $D$ meson decays (which is an excellent approximation in our setup), HFLAV directly provides constraints on the mixing amplitude parameterized by $x_{12} = 2 |M_{12}| \tau_D$, with the neutral $D$ meson lifetime $\tau_D \simeq 4.1 \times 10^{-13}$~s, and the CP violating phase $\phi_{12}$. We find that the confidence regions in the $x_{12} - \phi_{12}$ plane shown by HFLAV~\cite{HeavyFlavorAveragingGroup:2022wzx} can be reproduced with high accuracy by 
\begin{equation}
 x_{12} \cos(\phi_{12}) = (0.407 \pm 0.044)\% ~,\qquad x_{12} \sin(\phi_{12}) = (0.0045 \pm 0.0065)\% ~.
\end{equation}
The SM prediction for the $D$ mixing amplitude is expected to be real to a good approximation, but its size is not well known (see e.g.~\cite{Lenz:2020awd} for a review). In our numerical analysis, we allow the SM contribution to saturate the experimental central value of the real part with 100\% uncertainty
\begin{equation}
 M_{12}^\text{SM} = \frac{1}{2 \tau_D} \times (0.407 \pm 0.407)\% ~.
\end{equation}
To find bounds on the Higgs masses, we set the absolute value of the parameter combination $x_{cu}^* x_{uc}$, which enters the new physics contribution to the mixing amplitude, to 1 and marginalizing over its phase. We find
\begin{eqnarray}
    m_A \gtrsim \tan\beta \times 250 ~\text{GeV} ~,~~~ m_{A^\prime} \gtrsim \tan\beta^\prime \times 250 ~\text{GeV} ~, && ~~~ \text{from} ~D~ \text{mixing}~.
\end{eqnarray}
This is considerably weaker than the corresponding constraint from kaon mixing. We do not find a meaningful bound from $D$ mixing when we profile over the new physics phase. The constraints from $D$ mixing can be avoided entirely if the $x_{ij}$ parameters in the up sector are set to zero.

\bigskip
So far, we have seen that meson mixing generically puts very strong constraints on the model. In the presence of $\mathcal O(1)$ flavor violating parameters $x_{ij}$, $y_{ij}$ with $\mathcal O(1)$ CP violating phases, the strongest constraint comes from kaon mixing, and the results above suggest that the additional Higgs bosons have to have masses of at least $\sim$10~TeV or even $\sim$100~TeV for moderately large $\tan\beta$ and $\tan\beta^\prime$. 
\footnote{We arrived at these conclusions by considering the contributions to meson mixing from the tree-level exchange of neutral Higgs bosons. Additional contributions arise at the loop-level from box diagrams involving the charged Higgs bosons. While a detailed discussion of the charged Higgs boxes is beyond the scope of this paper, we have estimated the parametrically leading contributions to kaon mixing and find that they are very small. Relative to the tree-level contributions from the neutral Higgs bosons, they are suppressed by 
$$\frac{1}{16 \pi^2} ~\tan^2\beta~ \frac{m_c^2}{v^2} V_{cd}^* V_{cs} \quad \text{or} \quad \frac{1}{16 \pi^2} ~\tan^2\beta^\prime~\frac{m_u^2}{v^2} V_{ud}^* V_{us} ~,$$
for diagrams containing two $H^{\pm}$ bosons or two $H^{\prime \, \pm}$ bosons, respectively. The suppression by the loop factor and the small quark masses cannot be compensated for by reasonable values of $\tan\beta$ and $\tan\beta^\prime$. We therefore believe it is justified to neglect them in our analysis. We expect that similar conclusions hold for $B$ meson mixing.}

If the new flavor violating parameters are very small, $|x_{ij}|, |y_{ij}| \ll 1$, an irreducible amount of flavor mixing remains due to the CKM matrix. However, in that case, the new physics contributions to meson mixing turn out to be fairly small, and Higgs boson masses below the 1~TeV scale can be compatible with constraints from meson mixing.
Therefore, it is important to consider additional flavor observables that are sensitive to the new Higgs bosons and could provide stronger constraints.

\subsection{\texorpdfstring{\boldmath 
Rare $B$ meson decays $B_s \to \ell^+\ell^-$ and $B^0 \to \ell^+\ell^-$}{Rare B meson decays Bs -> l+l- and B0 -> l+l-}}

The rare decays of neutral $B$ mesons into a charged lepton pair, $B_s \to \ell^+\ell^-$ and $B^0 \to \ell^+\ell^-$, the $B_s \to \mu^+\mu^-$ decay in particular, are known to be sensitive probes of extended scalar sectors~\cite{Li:2014fea, Altmannshofer:2017wqy, Arnan:2017lxi}. 

The branching ratios of these $B$ meson decays can be predicted with very high precision. The largest uncertainty stems from the relevant CKM matrix elements. Using the results from~\cite{Bobeth:2013uxa, Beneke:2019slt} and the CKM input from eq.~\eqref{eq:CKM_input} we find 
\begin{eqnarray}
    \text{BR}(B_s\to \mu^+\mu^-)_\text{SM}
    &=&
    (3.51 \pm 0.22)\times 10^{-9}
    ~, \\
    \text{BR}(B^0\to \mu^+ \mu^-)_\text{SM}
    &=&
    (0.966 \pm 0.061) \times 10^{-10}
    ~, \\
    \text{BR}(B_s\to e^+e^-)_\text{SM}
    &=&
    (8.21 \pm 0.50)\times 10^{-14}
    ~, \\
    \text{BR}(B^0\to e^+ e^-)_\text{SM}
    &=&
    (2.26 \pm 0.14)\times 10^{-15}
    ~.
\end{eqnarray}
On the experimental side, $B_s\to\mu^+\mu^-$ is well established, while only upper bounds exist for $B^0\to\mu^+\mu^-$. The world averages from the PDG are based on LHCb, CMS, and ATLAS results~\cite{ATLAS:2018cur, LHCb:2021awg, CMS:2022mgd, ParticleDataGroup:2024cfk}, and read
\begin{eqnarray}
    \text{BR}(B_s\to\mu^+\mu^-)_\text{exp}
    &=&
    (3.34 \pm 0.27)\times 10^{-9}
    ~, \\
    \text{BR}(B^0\to\mu^+\mu^-)_\text{exp}
    &<&
    1.5 \times 10^{-10} ~~~ @ ~95\% ~ \text{C.L.}
    ~.
\end{eqnarray}
The experimental sensitivities to $B_s\to e^+e^-$ and $B^0 \to e^+e^-$ are still far above the SM predictions. The strongest constraints on the branching ratios come from LHCb~\cite{LHCb:2020pcv} 
\begin{eqnarray} \label{eq:Bsee_exp}
    \text{BR}(B_s\to e^+e^-)_\text{exp}
    &<&
    11.2 \times 10^{-9}  ~~~ @ ~95\% ~ \text{C.L.}
    ~, \\ \label{eq:B0ee_exp}
    \text{BR}(B^0\to e^+ e^-)_\text{exp}
    &<&
    3.0 \times 10^{-9}  ~~~ @ ~95\% ~ \text{C.L.}
    ~.
\end{eqnarray}
In the presence of NP, the expression for BR$(B_s\rightarrow\ell^+\ell^-)$ may be generally written as~\cite{DeBruyn:2012wk, Altmannshofer:2017wqy}
\begin{equation} \label{eq:Bsll}
    \frac{\text{BR}(B_s \to \ell^+\ell^-)}{\text{BR}(B_s \to \ell^+\ell^-)_{\text{SM}}}
    =
    \big(
    |S^{B_s}_{\ell\ell}|^2 + |P^{B_s}_{\ell\ell}|^2
    \big)
    \Big(
    \frac{1}{1 + y_s}
    +
    \frac{y_s}{y_s + 1}
    \frac{\text{Re}[(P^{B_s}_{\ell\ell})^2]-\text{Re}[(S^{B_s}_{\ell\ell})^2]}{|S^{B_s}_{\ell\ell}|^2 + |P^{B_s}_{\ell\ell}|^2}
    \Big)
    ~,
\end{equation}
where the effective lifetime difference in the $B_s$ meson system is parameterized by $y_s=(6.4 \pm 0.4)\%$ \cite{HeavyFlavorAveragingGroup:2022wzx}. 
In writing eq.~\eqref{eq:Bsll}, we have ignored a possible non-standard $B_s$ mixing phase $\phi_s$, which is justified given the strong constraints from measurements~\cite{HeavyFlavorAveragingGroup:2022wzx, LHCb:2023sim} (see Table~\ref{tab:mesonexp}).

The coefficients $S$ and $P$ depend on possible new physics contributions. In the SM, $P^{B_s}_{\ell\ell}=1$ and $S^{B_s}_{\ell\ell}=0$, while in our 3HDM setup we find
\begin{multline}
 S^{B_s}_{\ell\ell} = -\sqrt{1 - \frac{4 m_\ell^2}{m_{B_s}^2}} \frac{1}{C_{10}^\text{SM}} \frac{4\pi^2}{e^2} \Bigg( \sum_{i = h, H, H^\prime} \frac{m_{B_s}^2}{m_i^2} ~ \text{Re}(\kappa_{\ell\ell}^i) \left( \frac{m_s}{m_b} \kappa_{bs}^{i\,*}  - \kappa_{sb}^i \right) \frac{1}{V_{tb} V_{ts}^*} \\
 + \sum_{i = A, A^\prime} \frac{m_{B_s}^2}{m_i^2} ~ i \text{Im}(\kappa_{\ell\ell}^i) \left( \frac{m_s}{m_b} \kappa_{bs}^{i\,*}  + \kappa_{sb}^i \right) \frac{1}{V_{tb} V_{ts}^*} \Bigg) ~,
\end{multline}
\begin{multline}
 P^{B_s}_{\ell\ell} = 1 - \frac{1}{C_{10}^\text{SM}} \frac{4\pi^2}{e^2} \Bigg( \sum_{i = h, H, H^\prime} \frac{m_{B_s}^2}{m_i^2} ~ i\text{Im}(\kappa_{\ell\ell}^i) \left( \frac{m_s}{m_b} \kappa_{bs}^{i\,*}  - \kappa_{sb}^i \right) \frac{1}{V_{tb} V_{ts}^*} \\
 + \sum_{i = A, A^\prime} \frac{m_{B_s}^2}{m_i^2} ~ \text{Re}(\kappa_{\ell\ell}^i) \left( \frac{m_s}{m_b} \kappa_{bs}^{i\,*}  + \kappa_{sb}^i \right) \frac{1}{V_{tb} V_{ts}^*} \Bigg) ~.
\end{multline}
The SM Wilson coefficient is $C^{\text{SM}}_{10}\simeq-4.1$.
Analogous expressions hold also for $B^0 \to \ell^+ \ell^-$. Note that the width difference in the $B^0$ system is negligibly small $y_d \simeq 0$. 

To obtain an analytic understanding of the constraints that can be obtained from these rare decays, we expand the expressions in the limit $v^2 \ll m_A^2, m_{A^\prime}^2$ and $1 \ll \tan\beta \ll \tan\beta^\prime$. We also assume that the flavor-violating parameters are negligible $|x_{ij}|, |y_{ij}| \ll 1$. In that case, the irreducible contributions to $B_s \to \mu^+ \mu^-$ are given by
\begin{equation}
\label{Bstomumu}
S_{\mu\mu}^{B_s} \simeq - \frac{1}{C_{10}^\text{SM}} \frac{4 \pi^2}{e^2} \left( \frac{m_{B_s}^2}{m_A^2} + \frac{m_{B_s}^2}{m_{A^\prime}^2} \tan^2\gamma \right) \tan^2\beta ~,~~~ 
P_{\mu\mu}^{B_s} \simeq 1 - S_{\mu\mu}^{B_s} ~.
\end{equation}
As one can expect, the dominant contributions to $B_s \to \mu^+ \mu^-$ come from the ``second generation'' Higgs bosons. The contributions from the ``first generation'' Higgs bosons are suppressed by a factor $\tan^2\gamma \sim 1 / \tan^2\beta^\prime \ll 1$. Also note that $\tan\gamma$ is a free parameter and can in principle be made arbitrarily small. No robust constraints can therefore be obtained on $m_{A^\prime}$. 

A similar picture emerges for $B^0 \to \mu^+ \mu^-$. For this decay, we find
\begin{equation}
\label{B0tomumu}
S_{\mu\mu}^{B^0} \simeq \frac{1}{C_{10}^\text{SM}} \frac{4 \pi^2}{e^2} \left( \frac{m_{B^0}^2}{m_A^2} \frac{V_{cd}^*V_{cb}}{V_{td}^* V_{tb}} - \frac{m_{B^0}^2}{m_{A^\prime}^2} \frac{V_{ud}^*V_{ub}}{V_{td}^* V_{tb}} \frac{\tan\gamma\tan\beta^\prime}{\tan\beta} \right) \tan^2\beta ~,~~~
P_{\mu\mu}^{B^0} \simeq 1 - S_{\mu\mu}^{B^0} ~.
\end{equation}

Focusing on the ``second generation'' Higgs bosons, we estimate the following bounds 
\begin{eqnarray} \label{eq:Bsmm_constraint}
49~\text{GeV} \lesssim \frac{m_A}{\tan\beta} \lesssim 55~\text{GeV} ~~\text{or}~~ m_A \gtrsim \tan\beta \times 142~\text{GeV} ~, && ~~~ \text{from} ~B_s \to \mu^+ \mu^- ~, \\
m_A \gtrsim \tan\beta \times 50~\text{GeV} ~, && ~~~ \text{from} ~B^0 \to \mu^+ \mu^- ~.
\end{eqnarray}
The small low-mass window that is allowed by $B_s \to \mu^+\mu^-$ corresponds to a large new physics amplitude that interferes destructively with the SM and gives a SM-like branching ratio. Such a scenario is starting to be disfavored by measurements of the effective $B_s \to \mu^+ \mu^-$ lifetime~\cite{Altmannshofer:2017wqy, Altmannshofer:2021qrr}, but not fully excluded yet by $B_s \to \mu^+ \mu^-$ alone. 

For moderate $\tan\beta$, the bounds we find from $B_s \to \mu^+ \mu^-$ and $B^0 \to \mu^+ \mu^-$ are stronger than the ones from $B_s$ and $B^0$ mixing in eqs.~\eqref{eq:Bs_mixing_constraint} and \eqref{eq:B0_mixing_constraint}, and their strength increases with increasing $\tan\beta$. 

We checked if relevant bounds can be obtained from $B_s \to e^+ e^-$ and $B^0 \to e^+ e^-$. Using the same approximations as above, we find the new physics contributions
\begin{eqnarray}
\label{Bstoee}
S_{ee}^{B_s} &\simeq& - \frac{1}{C_{10}^\text{SM}} \frac{4 \pi^2}{e^2} \left( \frac{m_{B_s}^2}{m_A^2} - \frac{m_{B_s}^2}{m_{A^\prime}^2} \right) \tan\beta \tan\beta^\prime \tan\gamma ~, ~~~ P_{ee}^{B_s} \simeq 1 - S_{ee}^{B_s} ~,
\\
\label{B0toee}
S_{ee}^{B^0} &\simeq& \frac{1}{C_{10}^\text{SM}} \frac{4 \pi^2}{e^2} \left( \frac{m_{B^0}^2}{m_A^2} \frac{V_{cd}^* V_{cb}}{V_{td}^* V_{tb}} \frac{\tan\beta \tan\gamma}{\tan\beta^\prime} + \frac{m_{B^0}^2}{m_{A^\prime}^2} \frac{V_{ud}^* V_{ub}}{V_{td}^* V_{tb}} \right) \tan^2\beta^\prime  ~, ~~~ P_{ee}^{B^0} \simeq 1 - S_{ee}^{B^0} ~.
\end{eqnarray}

Due to the weak experimental constraints, see eqs.~\eqref{eq:Bsee_exp} and~\eqref{eq:B0ee_exp}, we do not obtain meaningful bounds on the Higgs bosons.

\subsection{\texorpdfstring{\boldmath Rare kaon decays $K^0 \to \ell^+ \ell^-$}{Rare kaon decays K0 -> l+l-}}

We also investigated the constraints that can be obtained from rare kaon decays. In particular, one can expect that the decay $K_L \to e^+ e^-$ can provide relevant constraints on the ``first generation'' Higgs bosons. In the following, we will discuss $K_L \to e^+ e^-$ and $K_L \to \mu^+ \mu^-$. The corresponding $K_S$ decays are not yet observed, and existing limits on their branching ratios~\cite{KLOE:2008acb, LHCb:2020ycd} have much weaker sensitivity to our scenario.

The $K_L \to \ell^+ \ell^-$ decays receive large long-distance contributions from $K_L \to \gamma \gamma$. Adapting the results from~\cite{Isidori:2003ts, Cirigliano:2011ny, Hoferichter:2023wiy}
we find the following expression for the $K_L \to \ell^+ \ell^-$ branching ratios in our 3HDM scenario
\begin{multline}
 \frac{\text{BR}(K_L \to \ell^+ \ell^-)}{\text{BR}(K_L \to \gamma\gamma)} = \frac{2 \alpha_\text{em}^2}{\pi^2} \frac{m_\ell^2}{m_K^2} \sqrt{1 - \frac{4m_\ell^2}{m_K^2}} \Bigg[ \big( \text{Im}\, C_{\gamma\gamma}\big)^2 + \\
 \Bigg( \text{Re}\, C_{\gamma\gamma} - \frac{5}{2} + \frac{3}{2} \log\left(\frac{m_\ell^2}{\mu^2}\right) + \chi_{\gamma\gamma}^{(\ell)}(\mu) + \chi_\text{SD}^\text{SM} + \chi_\text{SD}^P|_{\ell\ell} \Bigg)^2 + \big( \chi_\text{SD}^S|_{\ell\ell} \big)^2 \left(1 - \frac{4 m_\ell^2}{m_K^2} \right) \Bigg] ~.
\end{multline}
The imaginary and real parts of the coefficient $C_{\gamma\gamma}$ are given by
\begin{eqnarray} \label{eq:ImCgaga}
 \text{Im}\, C_{\gamma\gamma} &=& \frac{\pi}{2\beta_\ell} \log\left( \frac{1 - \beta_\ell}{1 + \beta_\ell} \right) \simeq \frac{\pi}{2} \log\left( \frac{m_\ell^2}{m_K^2}\right) ~, \\ \label{eq:ReCgaga}
 \text{Re}\, C_{\gamma\gamma} &=& \frac{1}{\beta_\ell} \Bigg[ \text{Li}_2\left(\frac{\beta_\ell - 1}{\beta_\ell + 1}\right) + \frac{\pi^2}{12} + \frac{1}{4}  \log^2\left( \frac{1 - \beta_\ell}{1 + \beta_\ell} \right) \Bigg] \simeq \frac{1}{4} \log^2\left(\frac{m_\ell^2}{m_K^2}\right) + \frac{\pi^2}{12}~,
\end{eqnarray}
where $\text{Li}_2(z) = -\int_0^z \frac{dx}{x} \log(1-x)$ is the di-logarithm function and $\beta_\ell = \sqrt{1 - 4 m_\ell^2/m_K^2}$ is the velocity of the leptons in the kaon restframe. The approximate expressions in equations~\eqref{eq:ImCgaga} and~\eqref{eq:ReCgaga} hold with high accuracy for electrons but not for muons.

The renormalization scale dependent function $\chi^{(\ell)}_{\gamma\gamma}(\mu)$ is a low energy coupling that is related to the off-shell $K_L \to \gamma \gamma$ form factor. Precise predictions have been obtained in~\cite{Hoferichter:2023wiy}, $\chi_{\gamma\gamma}^{(\mu)}(m_\rho) = 4.96 \pm 0.38$, $\chi_{\gamma\gamma}^{(e)}(m_\rho) = 8.0 \pm 1.0$ (see also the previous evaluations in~\cite{GomezDumm:1998gw, Isidori:2003ts, Cirigliano:2011ny}).

The short distance contributions to the $K_L \to \ell^+\ell^-$ decays are encoded in the SM coefficient $\chi_\text{SD}^\text{SM}$, and the lepton specific new physics contributions $\chi_\text{SD}^P|_{\ell\ell}$, and $\chi_\text{SD}^S|_{\ell\ell}$. 
Neglecting QED running effects and making use of isospin symmetry for the kaon masses $m_{K^0} \simeq m_{K^+} \equiv m_{K}$ and decay constants $f_{K^0} \simeq f_{K^+} \equiv f_K$, we find 
\begin{multline}
\sqrt{\frac{2 \tau_{K^+} \text{BR}(K_L \to \gamma\gamma)}{\tau_{K_L} \text{BR}(K^+ \to \mu^+ \nu)}} \left( 1 - \frac{m_\mu^2}{m_K^2} \right) \frac{m_\mu}{m_K} \big( \chi_\text{SD}^\text{SM} + \chi_\text{SD}^P|_{\ell\ell} \big) =  \frac{1}{s_W^2} \text{Re}\big( V_{td} V_{ts}^* Y(x_t) + V_{cd} V_{cs}^* Y_\text{NL} \big) \frac{1}{\lambda} \\
+ \frac{\pi}{\alpha_\text{em}} \Bigg[ \sum_{i = h,H,H^\prime} \frac{m_K^2}{m_i^2} \text{Im}\big( \kappa_{\ell\ell}^i\big) \text{Im}\left( \frac{m_d}{m_s} \kappa_{sd}^i - \kappa_{ds}^{i\,*}\right)\frac{1}{\lambda} \\ 
+ \sum_{i = A,A^\prime} \frac{m_K^2}{m_i^2} \text{Re}\big( \kappa_{\ell\ell}^i\big) \text{Re}\left( \frac{m_d}{m_s} \kappa_{sd}^i + \kappa_{ds}^{i\,*}\right) \frac{1}{\lambda} \Bigg] ~,
\end{multline}
\begin{multline}
\sqrt{\frac{2 \tau_{K^+} \text{BR}(K_L \to \gamma\gamma)}{\tau_{K_L} \text{BR}(K^+ \to \mu^+ \nu)}} \left( 1 - \frac{m_\mu^2}{m_K^2} \right) \frac{m_\mu}{m_K} ~ \chi_\text{SD}^S\big|_{\ell\ell} = \\
\frac{\pi}{\alpha_\text{em}} \Bigg[ \sum_{i = h,H,H^\prime} \frac{m_K^2}{m_i^2} \text{Re}\big( \kappa_{\ell\ell}^i\big) \text{Im}\left( \frac{m_d}{m_s} \kappa_{sd}^i - \kappa_{ds}^{i\,*}\right) \frac{1}{\lambda} \\ 
- \sum_{i = A,A^\prime} \frac{m_K^2}{m_i^2} \text{Im}\big( \kappa_{\ell\ell}^i\big) \text{Re}\left( \frac{m_d}{m_s} \kappa_{sd}^i + \kappa_{ds}^{i\,*}\right) \frac{1}{\lambda} \Bigg] ~,
\end{multline}
where $\lambda \simeq 0.225$~\cite{ParticleDataGroup:2024cfk} is the sine of the Cabibbo angle.
The SM part, $\chi_\text{SD}^\text{SM}$, contains the top contribution $Y(x_t) = 0.931\pm 0.005$~\cite{Brod:2022khx} and the charm contribution $Y_\text{NL} = \lambda^4 P_c = (2.84 \pm 0.26)\times 10^{-4}$~\cite{Gorbahn:2006bm, Hoferichter:2023wiy}. The $K_L \to \gamma \gamma$ and $K^+ \to \mu^+ \nu$ branching ratios and lifetimes that enter the above expressions are $\text{BR}(K_L \to \gamma \gamma) = (5.47 \pm 0.04) \times 10^{-4}$, $\text{BR}(K^+ \to \mu^+ \nu) = (63.56 \pm 0.11)\%$ and $\tau_{K_L} = (5.116 \pm 0.021) \times 10^{-8}$\,s, $\tau_{K^+} = (1.238 \pm 0.002)\times 10^{-8}$\,s~\cite{ParticleDataGroup:2024cfk}.

\begin{figure}[b]
\centering
\includegraphics[width = 1.0 \textwidth]{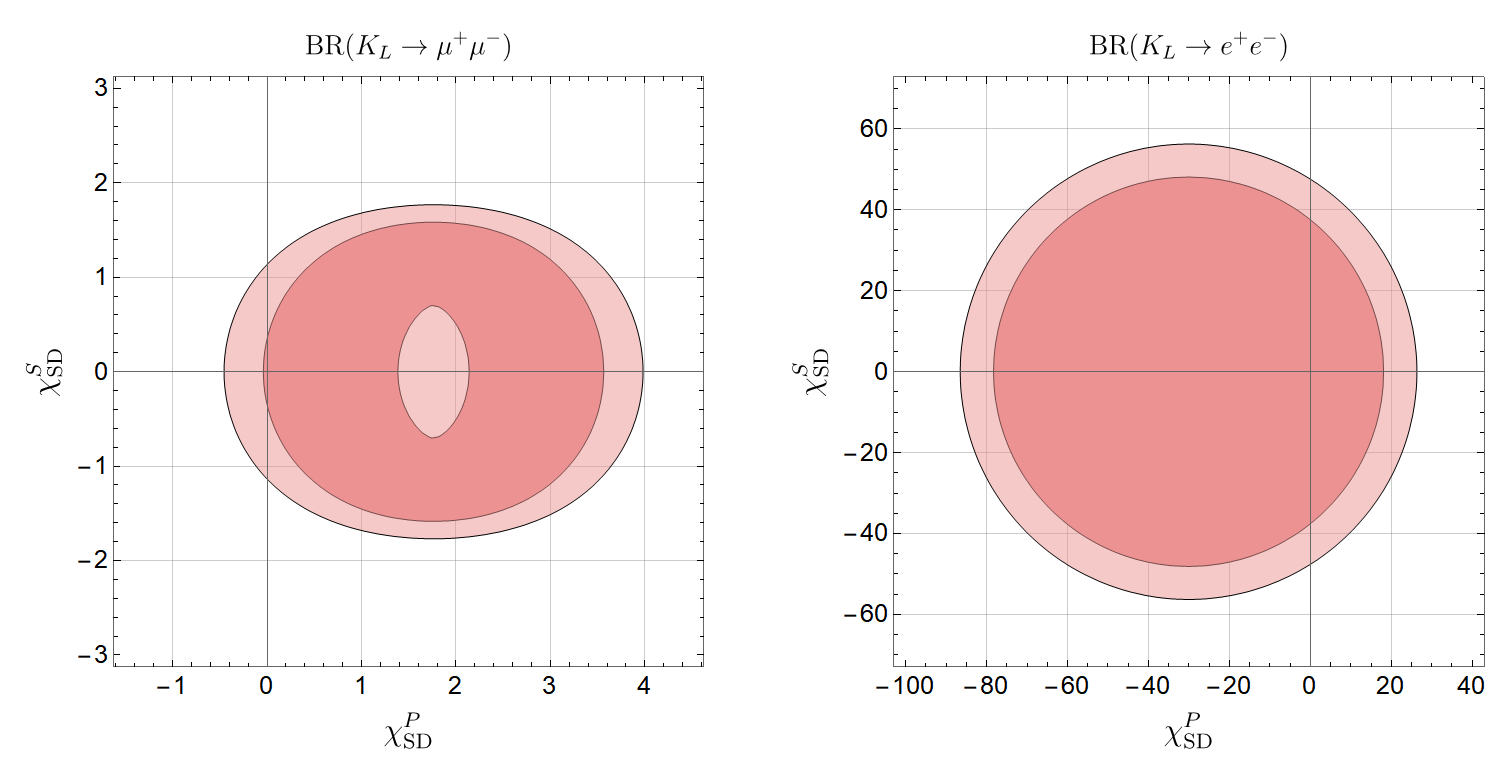}
\caption{Constraints on the new physics contributions $\chi_\text{SD}^P|_{\ell\ell}$ and $\chi_\text{SD}^S|_{\ell\ell}$ to the $K_L \to \ell^+ \ell^-$ decays. The plot on the left shows $K_L \to \mu^+ \mu^-$ and the plot on the right $K_L \to e^+ e^-$. The red shaded regions are allowed at the $1\sigma$ and $2\sigma$ level.}
\label{fig:KL_ll}
\end{figure}

Experimentally, the $K_L\to \mu^+ \mu^-$ decay is measured with high precision, while there is hardly evidence for the extremely rare $K_L\to e^+ e^-$ decay. Normalizing the corresponding branching ratios to the branching ratio of the $K_L \to \gamma \gamma$ decay one has~\cite{ParticleDataGroup:2024cfk, Hoferichter:2023wiy}
\begin{eqnarray}
\frac{\text{BR}(K_L \to \mu^+ \mu^-)}{\text{BR}(K_L \to \gamma \gamma)} \Bigg|_\text{exp} &=& (1.250 \pm 0.024) \times 10^{-5} ~, \\
\frac{\text{BR}(K_L \to e^+ e^-)}{\text{BR}(K_L \to \gamma \gamma)} \Bigg|_\text{exp} &=& (1.59^{+1.04}_{-0.75}) \times 10^{-8} ~.
\end{eqnarray}
Based on these experimental results, we find the allowed regions in $\chi_\text{SD}^P|_{\ell\ell}$ vs. $\chi_\text{SD}^S|_{\ell\ell}$ parameter space shown in figure~\ref{fig:KL_ll}.

In the phenomenologically interesting limit $v^2 \ll m_A^2, m_{A^\prime}^2$ and $1 \ll \tan\beta \ll \tan\beta^\prime$, and assuming that the flavor-violating parameters are negligible $|x_{ij}|, |y_{ij}| \ll 1$ we find the following approximate results for the new physics parameters $\chi_\text{SD}^P|_{\ell\ell}$ and $\chi_\text{SD}^S|_{\ell\ell}$ for muons
\begin{eqnarray}
\label{KLtomumu1}
\chi_\text{SD}^P\big|_{\mu\mu} &\simeq& - \sqrt{\frac{\tau_{K_L} \text{BR}(K^+ \to \mu^+ \nu)}{2 \tau_{K^+} \text{BR}(K_L \to \gamma\gamma)}} \left( 1 - \frac{m_\mu^2}{m_K^2} \right)^{-1} \frac{m_K}{m_\mu} \frac{\pi}{\alpha_\text{em}} \Bigg( \frac{m_K^2}{m_{A}^2} + \frac{m_K^2}{m_{A^\prime}^2} \frac{t_{\beta^\prime} t_\gamma}{t_\beta}\Bigg) t^2_\beta~, \\
\label{KLtomumu2}
\chi_\text{SD}^S\big|_{\mu\mu} &\simeq& 0 ~, 
\end{eqnarray}
and for electrons
\begin{eqnarray}
\label{KLtoee1}
\chi_\text{SD}^P\big|_{ee} &\simeq& \sqrt{\frac{\tau_{K_L} \text{BR}(K^+ \to \mu^+ \nu)}{2 \tau_{K^+} \text{BR}(K_L \to \gamma\gamma)}} \left( 1 - \frac{m_\mu^2}{m_K^2} \right)^{-1} \frac{m_K}{m_\mu} \frac{\pi}{\alpha_\text{em}} \Bigg( \frac{m_K^2}{m_{A^\prime}^2} - \frac{m_K^2}{m_{A}^2} \frac{t_{\beta} t_\gamma}{t_{\beta^\prime}}\Bigg) t^2_{\beta^\prime}~, \\
\label{KLtoee2}
\chi_\text{SD}^S\big|_{ee} &\simeq& 0 ~, 
\end{eqnarray}

As one might expect, we find relevant constraints on the second generation Higgs boson from $K_L \to \mu^+\mu^-$ and on the first generation Higgs boson from $K_L \to e^+e^-$. The approximate bounds on the Higgs masses are
\begin{eqnarray}
m_A \gtrsim \tan\beta \times 240~\text{GeV} ~, && ~~~ \text{from} ~K_L \to \mu^+ \mu^- ~, \\
m_A^\prime \gtrsim \tan\beta^\prime \times 31~\text{GeV} ~, && ~~~ \text{from} ~ K_L \to e^+ e^- ~.
\end{eqnarray}
The bound from $K_L \to \mu^+ \mu^-$ is particularly strong, surpassing the one from $B_s \to \mu^+\mu^-$ quoted in equation~\eqref{eq:Bsmm_constraint}.

\subsection{Summary of the flavor constraints} \label{sec:summary}

\begin{figure}[t]
\centering
\includegraphics[width = 1.0 \textwidth]{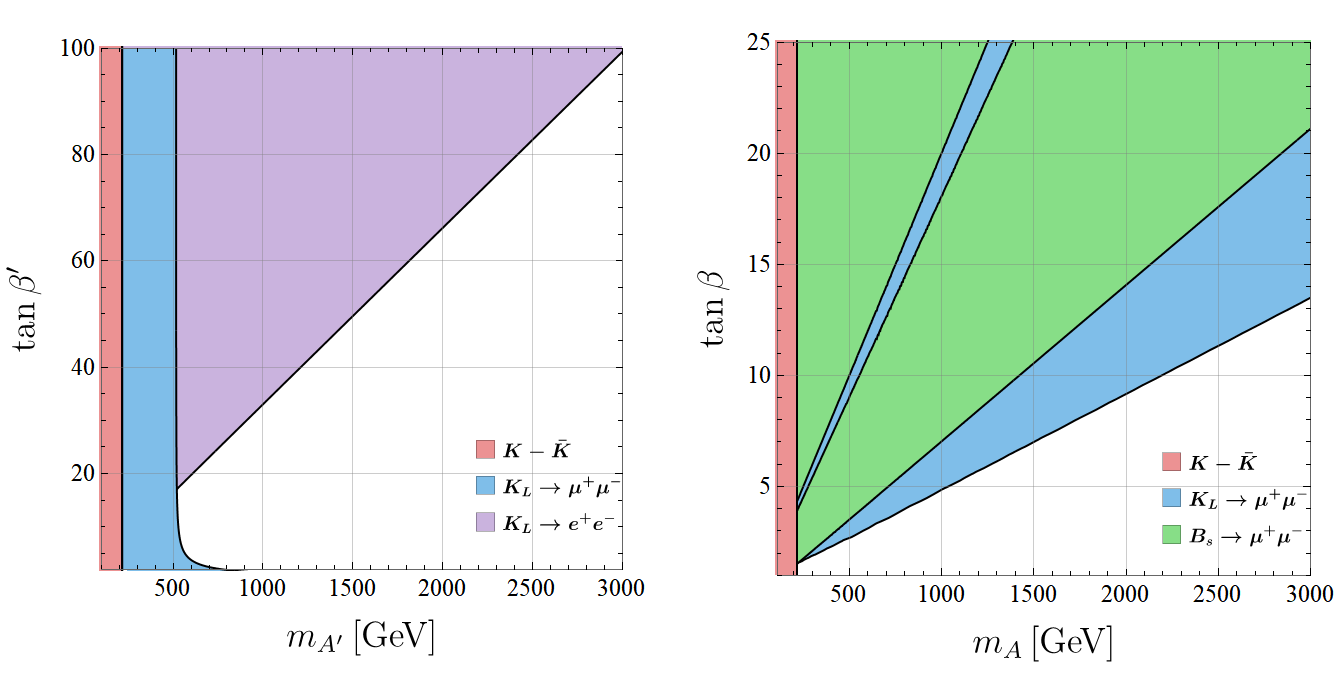}
\caption{Flavor constraints on the 3HDM parameter space in the generational limit, $x_{ij} , y_{ij}~=~0$. Left: constraints on the ``first generation'' Higgs boson parameters $m_{A^\prime}$ and $\tan\beta^\prime$. Right: constraints on the ``second generation'' Higgs boson parameters $m_A$ and $\tan\beta$. The colored regions are excluded at the 95\% C.L. by the indicated processes.}
\label{fig:constraints}
\end{figure}

As discussed in the previous section, neutral meson mixing constraints push the masses of the additional Higgs bosons far into the multi-TeV range if the flavor parameters $x_{ij}$ and $y_{ij}$ are generic (i.e. if their magnitudes are of order 1).
We therefore focus on the ``generation-specific'' limit, $x_{ij} , y_{ij} = 0$, which contains only the minimal amount of flavor violation to reproduce the observed quark mixing. 
In figure~\ref{fig:constraints}, we summarize the most important flavor constraints in this limit. The plot on the left shows the constraints on the ``first generation'' Higgs boson parameters $m_{A^\prime}$ and $\tan\beta^\prime$, while the one on the right shows the constraints on the ``second generation'' Higgs boson parameters $m_A$ and $\tan\beta$. The colored regions are excluded at the 95\% C.L. by the indicated processes. Only the most stringent constraints are shown.

To obtain the bounds displayed in figure~\ref{fig:constraints}, we use the generation specific relationships for each process within the context of the decoupling limit (i.e. equations~\eqref{KLtomumu1}-\eqref{KLtoee2} for the leptonic $K_L$ decays, equations~\eqref{Bstomumu}, \eqref{B0tomumu}, \eqref{Bstoee}, and \eqref{B0toee} for the leptonic $B^0$ and $B_s$ decays, and equations \eqref{B0mixing}, \eqref{Bsmixing}, and \eqref{Kmixing} for meson mixing). The $D$ meson mixing expression automatically vanishes in the ``generation specific" regime. We further fix $\tan\gamma = 1/\tan\beta^\prime$ and send the masses not being plotted to infinity. In other words, in obtaining constraints on the ``first generation" parameters of $m_{A^\prime}$ and $\tan\beta^\prime$, we send $m_A\rightarrow\infty$ and vice versa for the ``second generation " parameter space. This means that the shown bounds hold barring accidental cancellations among the contributions from different Higgs bosons. 

In the generation specific limit,
we find that $K_L\rightarrow e^+e^-$ provides the strongest bounds for the ``first-generation'' parameters $\tan\beta^\prime$ and $m_{A^\prime}$, whereas $K_L\rightarrow\mu^+\mu^-$ provides the strongest bounds for the ``second-generation'' parameters $\tan\beta$ and $m_{A}$, in particular for large $\tan\beta^\prime$ and $\tan\beta$ which is the best motivated region of parameter space. Kaon mixing is the most important constraint from meson mixing in both spaces, but is only relevant for low $\tan\beta^\prime$ and $\tan\beta$. As one might expect, the strongest constraints on the first (second) generation Higgs bosons come from observables in which the majority of the fermions in the initial and final states are from the first (second) generation.

For a moderate hierarchy in the vacuum expectation values of the 3HDM, $1 \ll \tan\beta \simeq 5 \ll \tan\beta^\prime \simeq 25$, the new Higgs boson masses can be comfortably as light as 1.5\,TeV without violating flavor constraints. For smaller values of $\tan\beta$ and $\tan\beta^\prime$, the Higgs bosons could be even lighter.

\section{Conclusions and Outlook}
\label{sec:conclusions}

In this paper, we explored a Three Higgs Doublet Model (3HDM) in which each of the three Higgs doublets primarily couples to a single generation of Standard Model fermions. One of the motivations for this scenario is its potential to partially address aspects of the SM flavor puzzle. In particular, the observed hierarchies in fermion masses could arise, at least in part, from a hierarchical pattern of Higgs vacuum expectation values, $v_1 \ll v_2 \ll v_3$. 

In the first part of the paper, we outlined the framework of our ``generational 3HDM'' in detail. The Yukawa sector is structured such that each Higgs doublet couples exclusively to one generation of fermions through rank-1 Yukawa matrices. A small number of free parameters governs the flavor misalignment among the three sets of up-type, down-type, and lepton Yukawa couplings. We assume that the flavor misalignment of the down-type Yukawas accounts for the observed CKM mixing in the SM quark sector, while the remaining flavor misalignment---parameterized by the coefficients $x_{ij}$ and $y_{ij}$---can, in principle, be set to zero. To accommodate the observed SM-like Higgs boson, we focused on the 3HDM alignment limit, which naturally emerges in the decoupling regime, where the additional Higgs bosons are significantly heavier than the electroweak scale. The key parameters governing the properties of the additional Higgs states are their masses, $m_A$ and $m_{A^\prime}$, as well as the ratios of vacuum expectation values $\tan\beta$ and $\tan\beta^\prime$, which characterize the ``second-generation Higgs'' (unprimed) and ``first-generation Higgs'' (primed), respectively.

Due to the non-trivial flavor structure of the model, the new Higgs bosons generically exhibit flavor-changing couplings. In the second part of the paper, we constrained the model's parameter space using flavor-changing neutral current processes. For generic flavor violation, where the parameters $x_{ij}$ and $y_{ij}$ are of order unity, we find that neutral meson mixing, particularly kaon mixing, imposes stringent constraints, pushing the new Higgs bosons well beyond the TeV scale and out of collider reach. However, in a scenario with the minimum amount of flavor violation (where the $x_{ij}$ and $y_{ij}$ are set to zero), the flavor constraints are significantly relaxed. In this case, the strongest bounds arise from rare leptonic decays of kaons and $B$ mesons, and Higgs bosons around the TeV scale remain viable. An interesting feature with regard to the flavor phenomenology is that the $m_{A^\prime}$ - $\tan\beta^\prime$ parameter space is primarily constrained by processes involving mainly first generation fermions (e.g. $K_L \to e^+e^-$), while the $m_A$ - $\tan\beta$ parameter space is constrained by processes involving mainly second generation fermions (e.g. $K_L \to \mu^+ \mu^-$ or $B_s \to \mu^+ \mu^-$). The most important bounds are summarized in Figure~\ref{fig:constraints}.

Our work motivates various follow up studies. Given that there are regions of parameter space in which the additional neutral Higgs bosons can have masses around the TeV scale, it would be interesting to explore the characteristic collider phenomenology. Based on the generational structure of the new Higgs boson couplings, one can expect that di-lepton resonance searches and di-jet resonance searches might already give relevant constraints on the model. Moreover, the flavor and collider signatures of the charged Higgs bosons warrant further investigation. The charged Higgs bosons can contribute to leptonic decays of charged mesons and might be probed by lepton flavor universality tests in $\pi^+ \to \ell^+ \nu$~\cite{PiENu:2015seu, PIONEER:2022yag} and $K^+ \to \ell^+ \nu$ decays~\cite{NA62:2012lny}. Finally, throughout much of our analysis, we have assumed that the tree-level Higgs potential respects CP invariance. It would be interesting to examine how our conclusions might change in the presence of CP violation in the Higgs sector.

\section*{Acknowledgments}

We thank Aditya Gadam and Stefania Gori for useful discussions. The research of W.A. and K.T. is supported by the U.S. Department of Energy grant number DE-SC0010107.

\begin{appendix}

\section{RGE running corrections to meson mixing} \label{appendix:mix}

In section~\ref{sec:mixing} we discussed constraints on the Higgs bosons from meson mixing observables and included the effect of RGE running from the scale of the Higgs bosons to the scale of the mesons. In this appendix we provide the relevant RGE factors and ratios of hadronic matrix elements.
We obtain the RGE factors from the 1-loop anomalous dimensions given in~\cite{Buras:2000if}. For $B$ meson mixing we find for the $X_i$ factors
\begin{multline}
    X_2(\mu) = -\frac{5}{16} \frac{B_2(m_b)}{B_1(m_b)} \Bigg[ \left( 1 - \frac{15}{\sqrt{241}} \right) \left( \frac{\alpha_s(\mu)}{\alpha_s(m_t)} \right)^{\frac{25+\sqrt{241}}{21}} \left( \frac{\alpha_s(m_t)}{\alpha_s(m_b)} \right)^{\frac{19+\sqrt{241}}{23}} \\ + \left( 1 + \frac{15}{\sqrt{241}} \right) \left( \frac{\alpha_s(\mu)}{\alpha_s(m_t)} \right)^{\frac{25-\sqrt{241}}{21}} \left( \frac{\alpha_s(m_t)}{\alpha_s(m_b)} \right)^{\frac{19-\sqrt{241}}{23}} \Bigg]~,
\end{multline}
\begin{multline}
    X_3(\mu) = \frac{1}{8 \sqrt{241}} \frac{B_3(m_b)}{B_1(m_b)} \Bigg[ \left( \frac{\alpha_s(\mu)}{\alpha_s(m_t)} \right)^{\frac{25+\sqrt{241}}{21}} \left( \frac{\alpha_s(m_t)}{\alpha_s(m_b)} \right)^{\frac{19+\sqrt{241}}{23}} \\ - \left( \frac{\alpha_s(\mu)}{\alpha_s(m_t)} \right)^{\frac{25-\sqrt{241}}{21}} \left( \frac{\alpha_s(m_t)}{\alpha_s(m_b)} \right)^{\frac{19-\sqrt{241}}{23}} \Bigg]~,
\end{multline}
\begin{equation}
    X_4(\mu) = \frac{3}{4} \left( 1 + \frac{m_b^2(m_b)}{6 m_{B_q}^2} \right) \frac{B_4(m_b)}{B_1(m_b)} \left( \frac{\alpha_s(m_t)}{\alpha_s(m_b)} \right)^{-\frac{6}{23}} ~,
\end{equation}
where we used the definition of the hadronic matrix elements from~\cite{FermilabLattice:2016ipl}.
The scale $\mu$ should be of the order of the heavy Higgs masses $m_H, m_A, m_{H^\prime}, m_{A^\prime}$. The ratios of the bag parameters $B_i(m_b)$ that enter the above expressions can be taken directly from table XV in~\cite{FermilabLattice:2016ipl}.

The RGE factors $\eta_i$ that are relevant for kaon and $D$ meson mixing are given by
\begin{multline}
\label{eq:eta2}
    \eta_2(\mu) =  \frac{1}{2} \Bigg[ \left( 1 - \frac{15}{\sqrt{241}} \right) \left( \frac{\alpha_s(\mu)}{\alpha_s(m_t)} \right)^{\frac{25+\sqrt{241}}{21}} \left( \frac{\alpha_s(m_t)}{\alpha_s(m_b)} \right)^{\frac{25+\sqrt{241}}{23}} \left( \frac{\alpha_s(m_b)}{\alpha_s(\mu_\text{low})} \right)^{\frac{25+\sqrt{241}}{25}} \\ + \left( 1 + \frac{15}{\sqrt{241}} \right) \left( \frac{\alpha_s(\mu)}{\alpha_s(m_t)} \right)^{\frac{25-\sqrt{241}}{21}} \left( \frac{\alpha_s(m_t)}{\alpha_s(m_b)} \right)^{\frac{19-\sqrt{241}}{23}} \left( \frac{\alpha_s(m_b)}{\alpha_s(\mu_\text{low})} \right)^{\frac{25-\sqrt{241}}{25}} \Bigg] ~,
\end{multline}
\begin{multline}
\label{eq:eta3}
    \eta_3(\mu) = \frac{1}{\sqrt{241}} \Bigg[ \left( \frac{\alpha_s(\mu)}{\alpha_s(m_t)} \right)^{\frac{25+\sqrt{241}}{21}} \left( \frac{\alpha_s(m_t)}{\alpha_s(m_b)} \right)^{\frac{25+\sqrt{241}}{23}} \left( \frac{\alpha_s(m_b)}{\alpha_s(\mu_\text{low})} \right)^{\frac{25+\sqrt{241}}{25}} \\ - \left( \frac{\alpha_s(\mu)}{\alpha_s(m_t)} \right)^{\frac{25-\sqrt{241}}{21}} \left( \frac{\alpha_s(m_t)}{\alpha_s(m_b)} \right)^{\frac{25-\sqrt{241}}{23}} \left( \frac{\alpha_s(m_b)}{\alpha_s(\mu_\text{low})} \right)^{\frac{25-\sqrt{241}}{25}} \Bigg]~,
\end{multline}
\begin{equation}
\label{eq:eta4}
    \eta_4(\mu) = 1 ~.
\end{equation}
As above, the scale $\mu$ should be of the order of the heavy Higgs masses. The scale $\mu_\text{low}$ is the scale at which the kaon or $D$ meson bag parameters are evaluated. In our numerical analysis we use the bag parameters from~\cite{Carrasco:2015pra}.

\section{A model for the rank-1 Yukawa couplings}
\label{app:rank1}

In this appendix we outline a simple construction that gives the rank-1 Yukawa couplings for the three Higgs doublets that we discussed in section~\ref{sec:yukawa}. If the three generations of SM fermions do not couple directly to the Higgs doublets but instead mix with three separate generations of vector-like fermions, rank-1 Yukawa couplings automatically arise, see e.g.~\cite{Kagan:1989fp}. 

We start by introducing three $U(1)$ symmetries that act separately on the the three Higgs doublets 
\begin{equation}
 \Phi_1 = (1,0,0) ~,\quad \Phi_2 = (0,1,0) ~,\quad \Phi_3 = (0,0,1) ~,
\end{equation}
while all SM fermions remain uncharged. For each $U(1)$, we introduce a single generation of heavy vector-like matter, charged under the corresponding $U(1)$ such that Yukawa couplings with a single Higgs are allowed
\begin{equation}
 \mathcal L \supset \sum_a \Big( \lambda_{U_a} \bar Q_{L_a} \tilde \Phi_a U_{R_a} +\lambda_{D_a} \bar Q_{L_a} \Phi_a D_{R_a} + \lambda_{L_a}\bar L_{L_a} \Phi_a E_{R_a} \Big) ~+~ \text{h.c.} ~,
\end{equation}
where the sum over $a = 1,2,3$ runs over the three Higgs doublets.
We assume that the $U(1)$ symmetries are softly broken by mass mixing between the vector-like fermions and the SM fermions. After integrating out the vector-like fermions, the effective Yukawa interactions of the SM fermions take the form
\begin{eqnarray}
\lambda_{u_1}^{ij} = \lambda_{U_1} \xi_{Q_1}^i \xi_{U_1}^j ~,\quad \lambda_{u_2}^{ij} = \lambda_{U_2} \xi_{Q_2}^i \xi_{U_2}^j ~,\quad \lambda_{u_3}^{ij} = \lambda_{U_3} \xi_{Q_3}^i \xi_{U_3}^j ~, \\
\lambda_{d_1}^{ij} = \lambda_{D_1} \xi_{Q_1}^i \xi_{D_1}^j ~,\quad \lambda_{d_2}^{ij} = \lambda_{D_2} \xi_{Q_2}^i \xi_{D_2}^j ~,\quad \lambda_{d_3}^{ij} = \lambda_{D_3} \xi_{Q_3}^i \xi_{D_3}^j ~, \\
\lambda_{\ell_1}^{ij} = \lambda_{L_1} \xi_{L_1}^i \xi_{E_1}^j ~,\quad \lambda_{\ell_2}^{ij} = \lambda_{L_2} \xi_{L_2}^i \xi_{E_2}^j ~,\quad \lambda_{\ell_3}^{ij} = \lambda_{L_3} \xi_{L_3}^i \xi_{E_3}^j ~.
\end{eqnarray}
These Yukawa interactions are outer products of two flavor vectors and thus rank-1 by construction. 

As a by-product, the three $U(1)$ symmetries also greatly reduce the number of Higgs potential terms as discussed in section~\ref{sec:lagrangian}.

\end{appendix}

\bibliographystyle{jhep}
\bibliography{refs} 

\end{document}